\documentclass[useAMS,usenatbib,fleqn]{mn2e}

\usepackage{color} 
\usepackage{url} 
\usepackage{subfig}
\usepackage{graphicx}
\usepackage{hyperref}

\usepackage{amssymb} 

\usepackage{amsmath}
\DeclareMathOperator{\sgn}{sgn}

\usepackage{upgreek}

\usepackage{txfonts} 

\bibpunct{(}{)}{,}{a}{}{,} 

\title[]{Comparative performance of selected variability detection techniques
in photometric time series}
\author[]{\parbox{\textwidth}{K.~V.~Sokolovsky$^{1,2,3}$\thanks{E-mail: kirx@kirx.net},
P.~Gavras$^{1}$,
A.~Karampelas$^{1}$,
S.~V.~Antipin$^{2,4}$,
I.~Bellas-Velidis$^{1}$,
P.~Benni$^{5}$,
A.~Z.~Bonanos$^{1}$,
A.~Y.~Burdanov$^{6,7}$,
S.~Derlopa$^{8}$,
D.~Hatzidimitriou$^{1,9}$,
A.~D.~Khokhryakova$^{6}$,
D.~M.~Kolesnikova$^{4}$,
S.~A.~Korotkiy$^{10}$,
E.~G.~Lapukhin$^{11}$,
M.~I.~Moretti$^{1}$,
A.~A.~Popov$^{6}$,
E.~Pouliasis$^{1,9}$,
N.~N.~Samus$^{4,2}$,
Z.~Spetsieri$^{1,9}$,
S.~A.~Veselkov$^{11}$,
K.~V.~Volkov$^{6}$,
M.~Yang$^{1}$,
A.~M.~Zubareva$^{4,2}$}\vspace{0.4cm}\\
\parbox{\textwidth}{$^{1}$IAASARS, National Observatory of Athens, 15236 Penteli, Greece\\
$^{2}$Sternberg Astronomical Institute, Moscow State University, Universitetskii~pr. 13, 119992 Moscow, Russia\\
$^{3}$Astro Space Center of Lebedev Physical Institute, Profsoyuznaya Str. 84/32, 117997 Moscow, Russia\\
$^{4}$Institute of Astronomy (Russian Academy of Sciences), Pyatnitskaya Str. 48, 119017 Moscow, Russia\\
$^{5}$Acton Sky Portal, 3 Concetta Circle, Acton, MA 01720, USA\\
$^{6}$Kourovka Astronomical Observatory of Ural Federal University, Mira Str. 19, 620002 Ekaterinburg, Russia\\
$^{7}$Institut d'Astrophysique et G\'{e}ophysique, Universit\'{e} de Li\`{e}ge, all\'{e}e du 6 Ao\^{u}t 17, 4000 Li\`{e}ge, Belgium\\
$^{8}$Department of Physics, University of Patras, 26500 Patra, Greece\\
$^{9}$Department of Astrophysics, Astronomy \& Mechanics, Faculty of Physics, University of Athens, 15783 Athens, Greece\\
$^{10}$Ka-Dar astronomy foundation, Kuzminki, P.O. Box 82, 142717 Moscow region, Russia\\
$^{11}$Reshetnev Siberian State Aerospace University, Krasnoyarsky Rabochy~Av.\,31, 660037 Krasnoyarsk, Russia
}}

\begin{document}

\date{Accepted 2016 September 5. Received 2016 September 5; in original form 2016 May 2}

\pagerange{\pageref{firstpage}--\pageref{lastpage}} \pubyear{2016}

\maketitle

\label{firstpage}

\begin{abstract}
Photometric measurements are prone to systematic errors
presenting a challenge to low-amplitude variability detection.
In search for a general-purpose variability detection technique able
to recover a broad range of variability types including currently unknown ones,
we test 18 statistical characteristics quantifying 
scatter and/or correlation between brightness measurements.
We compare their performance in identifying variable objects in seven
time series data sets obtained with telescopes ranging in size from a
telephoto lens to 1\,m-class 
and probing variability on time-scales from minutes to decades.
The test data sets together include lightcurves of 127539 objects, among them 1251 variable stars of various types 
and represent a range of observing conditions
often found in ground-based variability surveys.
The real data are complemented by simulations.
We propose a combination of two indices that together
recover a broad range of variability types from photometric data
characterized by a wide variety of sampling patterns, photometric
accuracies and percentages of outlier measurements. The first index is the
interquartile range (IQR) of magnitude measurements, 
sensitive to variability irrespective of 
a time-scale and 
resistant to outliers. It
can be complemented by
the ratio of the lightcurve variance to the mean square successive
difference, $1/\eta$, 
which is efficient in detecting variability on time-scales longer than 
the typical time interval between observations.
Variable objects have larger $1/\eta$ and/or IQR values than
non-variable objects of similar brightness.
Another approach to variability detection is to
combine many variability indices using principal component analysis. 
We present 124 previously unknown variable stars found in the test data.
\end{abstract}


\section{Introduction}

A variety of phenomena manifest themselves as changes in
apparent brightness of astronomical objects. The amplitudes and time-scales of
these changes vary from tens of magnitudes and weeks for supernova
explosions to a fraction of a magnitude and minutes for stellar pulsations.
With the notable exceptions of light echoes
\citep[e.g.][]{2003Natur.422..405B}, variable reflecting nebulae
\citep[e.g.][]{1997ApJ...489..210C} and 
the M87 jet \citep[e.g.][]{2011ApJ...743..119P} variable objects
are unresolved by single-dish 
telescopes\footnote{The light travel time argument implies that
an object varying on a time-scale $t$ cannot be larger than $ct$ light
seconds, otherwise its variability would be smeared.}. 
Variable point-like objects are often embedded in light of a resolved 
non-variable source (active
nucleus or a supernova in a galaxy; young stellar object embedded in a
nebula) that complicate measurements of the variable object's brightness.
The variations may be associated with a single catastrophic event
(supernova), may be approximately (dwarf novae) or strictly periodic (eclipsing binaries) 
or aperiodic (active galactic nuclei) in nature. Our understanding of these events depends on 
the efficient and reliable detection of brightness variations.

Photometric measurements are prone to systematic errors that are difficult
to characterize. This makes it challenging to distinguish true low-amplitude
variability from the apparent one caused by systematic effects and
measurement errors.
Imaging artefacts such as cosmetic defects of a CCD, diffraction spikes from
bright objects and cosmic ray hits as well as blending between images of
nearby objects can corrupt photometry and mimic high-amplitude variability.
Three different lines of attack on the problem of variable object detection 
are described in the literature: direct image comparison, (``transient detection''), 
lightcurve analysis using variability indices and periodicity search.

Transient detection techniques seek to identify changes between two sets of
sky images taken at different times (epochs). The changes may be found by
subtracting the images pixel-by-pixel after resampling them to a common
coordinate grid and accounting for seeing changes 
(difference image analysis -- DIA;
\citealt{1996AJ....112.2872T}, 
\citealt{1998ApJ...503..325A,2000A&AS..144..363A},
\citealt{2008MNRAS.386L..77B,2012MNRAS.425.1341B},
\citealt{2015arXiv151206872Z}, \citealt{2016MNRAS.457..542B}; applications of the method include
\citealt{2003AJ....126..175B,2003AstL...29..599Z,2013AcA....63..429A,2014PASA...31...12S,2015RAA....15..215Z}).
Large surveys such as OGLE \citep{2015AcA....65....1U}, PTF~\citep{2009PASP..121.1395L}, 
Pan-STARRS \citep{2014ApJ...795...44R} and DES~\citep{2015AJ....150..172K} implement the image subtraction technique.
Alternatively, one may extract astronomical objects (sources) from each image independently and
compare the resulting source lists 
(\citealt{2014MNRAS.439.1829C}, CRTS -- \citealt{2009ApJ...696..870D}).
The second-epoch images are often taken in pairs, triplets or even longer series with dithering to eliminate 
image artefacts that are usually associated with a given position on the
image detector, not in the sky.

More sophisticated detection strategies may be applied if measurements are
obtained at more than two epochs. Their obvious advantage over the simple 
two-epoch data comparison is the potential to average-out
individual measurement errors and thus detect variability with a lower amplitude.
One class of methods employs various ``variability indices'' characterizing the overall scatter of measurements
in a lightcurve and/or degree of correlation between consecutive flux measurements
\citep[some recent examples:][see the detailed discussion in
Section~\ref{sec:indices}]{2014JAD....20....4M,2015MNRAS.447.3973J,2015AJ....150..107Y}.
The other class of methods 
search for
significant
periodicity in flux variations
\citep[e.g.][]{2014MNRAS.438.3383M,2014ApJS..213....9D,2014AcA....64..309K,2015ApJ...802L...4C,2015MNRAS.447.3536N,2016MNRAS.455.2337N,2015AcA....65..297S}.
While many types of variable stars show periodic or semi-periodic light
variations, flux measurement errors are expected to be aperiodic, or
associated with a known periodic process inherent to the observations
(diurnal or seasonal cycle, synodic month, periodic guiding errors, orbital period of a spaceborne
telescope, etc.).

If a search is aimed at a specific variability type for which a lightcurve
shape is generally known in advance (e.g. exoplanet transits or eclipsing
binaries in general, Cepheids, RR~Lyrae stars, novae), 
template fitting 
(e.g. \citealt{1996Icar..119..244J}, \citealt{1999ApJ...521..155M}, 
\citealt{2011AJ....141...83P,2013AJ....146...21S,2014A&A...567A.100A}, 
\citealt{2016arXiv160708658H}) 
with various trial periods/flare development time-scales can be performed.
Simple cuts on lightcurve parameters \citep{2008A&A...477...67H,2010AcA....60..109G} as well as
advanced machine learning techniques \citep{2005AJ....130...84F} can be used to
select lightcurves of a known shape from a large photometric data set.
A pre-selection based on colour can be used to reduce the number of candidates
when searching for variables of a specific type
\citep[][]{2006AJ....132.1202K,2013A&A...551A..77T,2014ApJ...781...22Z,2016MNRAS.455.2163O,2016MNRAS.459.1687M}.

Since period search and template-fitting algorithms are computationally expensive, a two-step
approach can be applied. Candidate variable stars are pre-selected using 
a fast-to-compute variability index (and/or colour) and only the lightcurves that passed this
selection are subjected to period search
\citep[e.g.][]{2000AJ....119.1901A,2013ApJ...763...32D,2014A&A...562A.125K,2015A&A...573A.100F,2015A&A...574A..15F,2015A&A...575A.114G,2016AJ....151..118V}
or template fitting \citep[e.g.][]{2011ApJ...733..124S,2015AJ....149..183H}.
If the total number of observed objects is low, both period search and
lightcurve scatter-based selection criteria are applied independently of each other to
conduct exhaustive search for both periodic and non-periodic variables 
\citep{2014AcA....64..115S,2015AJ....150..175R}.
Selection based on period search may be followed by 
even more computationally intensive steps like binary system modelling
\citep{2005ApJ...628..411D}.
\cite{2014A&A...566A..43K} used the period along with other variability
features as an input for the random forest algorithm to select periodic
variable star candidates in the EROS-2 database and simultaneously classify them.

The methods described above may efficiently select variable object candidates
from a large set of photometric data. However, the final decision to
designate an object as ``variable star'' rather than a ``candidate'' is usually made only
after visual inspection of the object's lightcurve by a human expert
(e.g.
\citealt{2005AcA....55..275P,2011AcA....61..103G,2013AcA....63..115P,2013AcA....63..323P,2013AJ....146..101P},
\citealt{2013ApJ...779....7C}, 
\citealt{2016AJ....151..110K},
\citealt{2016arXiv160604792S}).
If the number of observations is small, the original images are checked for
the presence of obvious problems [image artefacts, cosmic ray hits, 
point spread function (PSF) wings of a bright nearby object] affecting photometry of a candidate
variable
\citep[e.g.][]{2003AJ....125.1261D,2010ApJ...712.1259B,2011AstL...37...91D,2014MNRAS.437..132R}.
While advanced image artifact rejection procedures exist
\citep{2002PASP..114..144F,2016A&C....16...67D},
visual image inspection remains an important data quality control
tool as it may uncover unexpected problems \citep{2016A&C....16...99M}.

Variable star detection may be considered in the framework of classical
hypothesis testing \citep[e.g.][]{2003psa..book.....W}: 
to establish that an object is variable, one needs to
rule~out the null hypothesis that it is constant given the observations 
\citep{2006ASPC..349...15E}.
One may compare a value of variability-detection statistic
(Section~\ref{sec:indices}) derived from the lightcurve to the distribution of 
this value expected for non-variable objects. The problem is that objects 
with corrupted measurements produce long tails in the aforementioned distributions. 
In the presence of badly measured objects one is forced to set a low threshold for
accepting candidate variables (Section~\ref{sec:cutoffvalue}) and rely on
additional information not captured by the variability-detection statistic
to distinguish true variables from badly measured objects in the
distribution tail.

Alternatively one may view the search for variable stars as a classification problem
that may be approached with machine learning techniques.
The task is to classify a set of objects characterized by their lightcurves, 
images associated with each lightcurve point and possibly additional pieces of
information associated with each brightness measurement (object's position on the CCD
frame, airmass, seeing, temperature, etc.). 
One needs to distinguish
various classes of variable stars from the class of well-measured constant stars and
classes of stars affected by various types of measurement errors (bad pixels,
diffraction spikes, blending). Objects that do not belong to one of
the known classes should also be identified.
While considerable progress has been made in 
lightcurve-based automated classification of stars already known to be
variable \citep{2007A&A...475.1159D,2014AJ....148...31P,2016A&A...587A..18K}, 
an automated system that could reliably identify variable stars among
non-variables remains to be developed.

In practice, the following approach to variable star
detection is often adopted.
{(i)\,}Objects affected by blending and image artefacts 
are flagged at source extraction stage. 
{(ii)\,}The lightcurves of the detected 
objects are constructed and may be refined using the available additional information
(Section~\ref{sec:filtering}). 
{(iii)\,}The techniques described in the previous 
paragraphs are used to select promising variable star candidates based on their 
lightcurves. 
{(iv)\,}The list of candidates is examined by a human expert who
performs the final classification and removes false variables from the list.
In this work we explore the limits of the traditional approach outlined
above and identify the best ways to select candidate variables.

We compare the performance of popular variability detection techniques on various 
real and simulated photometric data sets.
We refer to any value that quantifies `how variable' a given object is as a `variability index'.
The discussion is limited to variability indices based on lightcurve
scatter (Sections~\ref{sec:chi2}--\ref{sec:peaktopeak}) and correlation
(Sections~\ref{sec:lag1autocorr}--\ref{sec:sb}) while the period-search based 
techniques will be discussed elsewhere.
We attempt to find a general-purpose variability detection technique able 
to recover a broad range of variability types including currently unknown
ones \citep{2009MNRAS.400.1897S}.
Such a technique would also be useful for solving the opposite problem:
reliable selection of non-variable objects that can be used as photometric
standards \citep[e.g.][]{2012PASP..124..854O} or targets for searches of
variations not intrinsic or not typical to the objects such as microlensing
events \citep{1994AcA....44..227U}, occultations of stars by distant Solar
system bodies \citep{2013AJ....146...14Z}, tidal disruption events in
nuclei of non-active galaxies \citep{2011ApJ...741...73V} 
and failed supernovae \citep{2008ApJ...684.1336K}.

Publications focused on comparing performance of variability search techniques include
\cite{2012A&A...548A..48E} who compared 
planetary-transit
detection algorithms, while
\cite{2010AJ....139.1269D} and \cite{2010ApJ...723..737V} discussed a number of
variability detection tests in the context of active galactic nuclei.
\cite{2016A&A...586A..36F} compared performance of some 
multi-band correlation-based variability indices.
\cite{2003MNRAS.345.1271V} and \cite{2013ApJ...771....9A} discussed properties 
of `excess variance'
(Section~\ref{sec:sigmaxs}) and `fractional variability amplitude', 
the variability measures often used in X-ray astronomy.
\cite{2013MNRAS.434.3423G} compared the accuracy and performance of period
finding algorithms.
\cite{2015ApJ...798...89F} compare various methods of extracting 
a characteristic time-scale from an irregular lightcurve. 
\cite{2015arXiv150600010N} provide an extensive list of features
useful for lightcurve characterization and classification.
Preliminary results of our variability index comparison based solely on 
photographic lightcurves are presented by \cite{2016arXiv160503571S}.

This paper is structured as follows.
Section~\ref{sec:indices} defines the variability indices we investigate.  
Section~\ref{sec:testdata} describes the test data. 
Section~\ref{sec:comparison_tech} presents the technique for comparison of
effectiveness of variability indices in selecting variable objects.
Section~\ref{sec:resdisc} discusses the results of the comparison 
and Section~\ref{sec:conclusions} summarizes our findings.

\section{Variability search methods}
\label{sec:indices}

In this section we define the numerical parameters characterizing 
the `degree of variability' of an object -- the variability indices,
discussed in detail in the following paragraphs. 
The scatter-based indices
(Sections~\ref{sec:sigma}--\ref{sec:peaktopeak}) consider only the distribution
of measured magnitudes ignoring the time information available in a
lightcurve. Some also take into account the estimated errors.
The correlation-based indices (Sections~\ref{sec:lag1autocorr}--\ref{sec:sb}) 
in addition to the measured magnitudes 
themselves consider the order in which the measurements were taken and 
some indices also take into account the time difference between measurements.
The use of this additional information makes correlation-based indices more
sensitive to low-amplitude variability, but on the downside, correlation-based indices are insensitive to variability on
time-scales shorter than the sampling time \citep{2011ApJ...735...68K}.
Table~\ref{tab:indexsummary} summarizes the information used by each index.
In the following sections we compare the effectiveness of these
variability indices in selecting variable stars.

\subsection{$\chi^2$ test}
\label{sec:chi2}

A $\chi^2$ test\footnote{\url{https://en.wikipedia.org/wiki/Chi-squared_test}}
is any statistical hypothesis test in which the sampling
distribution of the test statistic is a $\chi^2$ distribution when the
null hypothesis is true.
Given $N$ magnitude measurements $m_i$ (assumed to be independent of each
other) and their associated errors $\sigma_i$ (assumed to be Gaussian),
the null hypothesis, $H_0$, that an object does not change its brightness can be tested 
by computing the value
\begin{equation}
\chi^2 = \sum\limits_{i=1}^N \frac{(m_i-\bar{m})^2}{\sigma_i^2},
\end{equation}
where 
\begin{equation}
\label{eq:weighted_average}
\bar{m} = \sum\limits_{i=1}^N \frac{ m_i }{ \sigma_i^2 } / \sum\limits_{i=1}^N \frac{ 1 }{ \sigma_i^2 }
\end{equation}
is the weighted mean magnitude. 
$\chi^2$ is compared to the critical value $\chi_{p,\nu}^2$ obtained
from the $\chi^2$-distribution with $\nu = N - 1$ degrees of freedom.
The $p$-value indicates the statistical significance level at which $H_0$ can be
rejected.

If measurement errors are estimated correctly, the majority of objects
should have $\chi^2$ values consistent with $H_0$, since the majority of
stars are not variable. A notable exception from this rule is
millimagnitude-precision photometric observations such as the ones
obtained by MOST \citep{2003PASP..115.1023W}, CoRoT
\citep{2009A&A...506..411A}, Kepler \citep{2010Sci...327..977B} 
and future photometric space missions
\citep[e.g.][]{2014SPIE.9143E..20R,2014ExA....38..249R}, 
which are able to detect variability in the majority of field stars,
including variability caused by transiting Solar~system-like planets
\citep{2015ApJ...810...29H}.

In practice, poor knowledge of $\sigma_i$ limits the applicability of the $\chi^2$
test for variability detection in ground-based photometry.
In this case, $\chi^2$ may still be useful as a measure of scatter in a lightcurve 
compared to the expected measurement errors, but the cut-off value for discriminating 
variable objects from non-variable ones should be different from the one
suggested by the $\chi^2$ distribution.
In the following we use the reduced $\chi_{\rm red}^2 = \chi^2 / N - 1$
\citep[e.g.][]{2010arXiv1012.3754A}
to compare its value for lightcurves with different $N$.
\cite{2010ApJ...723..737V} note that
estimated photometric measurement errors are asymmetric and non-Gaussian when
converted from flux to magnitude space. This violates the assumptions behind
the critical value $\chi_{p,\nu}^2$ calculations. The $\chi^2$ test, in its
textbook form, should be performed in flux space 
and only when the contribution from all sources of photometric errors
has been properly accounted for.

\begin{table}
   \caption{Information included in variability indices. For references see the~footnote in Table~\ref{tab:FRreal}.}
   \label{tab:indexsummary}
   \begin{tabular}{c c@{~~}c@{~~}c c@{~~}c}
   \hline\hline
Index                     & Errors & Order & Time  & Sec. & Ref. \\
   \hline
\multicolumn{6}{c}{Scatter-based indices} \\
$\chi_{\rm red}^2$        & $\checkmark$ &              &              & \ref{sec:chi2}            & (a) \\
$\sigma$                  &              &              &              & \ref{sec:sigma}           & (b) \\
$\sigma_w$                & $\checkmark$ &              &              & \ref{sec:sigma}           & (b) \\
${\rm MAD}$               &              &              &              & \ref{sec:MAD}             & (c) \\
${\rm IQR}$               &              &              &              & \ref{sec:IQR}             & (d) \\
${\rm RoMS}$              & $\checkmark$ &              &              & \ref{sec:RoMS}            & (e) \\
$\sigma_{\rm NXS}^2$      & $\checkmark$ &              &              & \ref{sec:sigmaxs}         & (f) \\
$v$                       & $\checkmark$ &              &              & \ref{sec:peaktopeak}      & (g) \\
\multicolumn{6}{c}{Correlation-based indices} \\
$l_1$                     &              & $\checkmark$ &              & \ref{sec:lag1autocorr}    & (h) \\
$I$                       & $\checkmark$ & $\checkmark$ & $\checkmark$ & \ref{sec:welch}           & (i) \\
$J$                       & $\checkmark$ & $\checkmark$ & $\checkmark$ & \ref{sec:stetson}         & (j) \\
$J({\rm time})$           & $\checkmark$ & $\checkmark$ & $\checkmark$ & \ref{sec:stetsonTime}     & (k) \\
$J({\rm clip})$           & $\checkmark$ & $\checkmark$ & $\checkmark$ & \ref{sec:stetsonClip}     & (d) \\
$L$                       & $\checkmark$ & $\checkmark$ & $\checkmark$ & \ref{sec:stetson}         & (j) \\
${\rm CSSD}$              &              & $\checkmark$ &              & \ref{sec:cssd}            & (l) \\
$E_x$                     & $\checkmark$ & $\checkmark$ & $\checkmark$ & \ref{sec:excursions}      & (m) \\
$1/\eta$                  &              & $\checkmark$ &              & \ref{sec:vonneumann}      & (n) \\
$\mathcal{E}_\mathcal{A}$ &              & $\checkmark$ & $\checkmark$ & \ref{sec:EA} & (o) \\
$S_B$                     & $\checkmark$ & $\checkmark$ &              & \ref{sec:sb}              & (p) \\
   \hline
   \end{tabular}
   \renewcommand{\arraystretch}{1.0}
\end{table}

\subsection{Standard deviation, $\sigma_w$}
\label{sec:sigma}

A detectable variable star, by definition, should have larger scatter of
measurements in its lightcurve compared to non-variable stars that could be
measured with the same accuracy. One way to characterize scatter of
measurements is to compute the standard deviation,
\begin{equation}
\label{eq:sigma}
\sigma = \sqrt{ \frac{1}{N-1} \sum\limits_{i=1}^N (m_i-\bar{m})^2 }
\end{equation}
or alternatively, if the estimated errors are assumed to reflect the relative
accuracy of measurements, its weighted version
\begin{equation}
\sigma_w = \sqrt{ \frac{\sum\limits_{i=1}^N w_i}{(\sum\limits_{i=1}^N w_i)^2 - \sum\limits_{i=1}^N (w_i^2)} 
                \sum\limits_{i=1}^N w_i (m_i - \bar{m})^2 }
\end{equation}
Assuming that $m_i$ are drawn from Gaussian distributions having 
variances $\sigma_i^2$
and
the same mean $\bar{m}$, the choice of weights 
$w_i = 1/\sigma_i^2$ maximizes the likelihood of obtaining 
the set of measurements ($m_i$).
Therefore, given a set of measurements ($m_i$, $\sigma_i$), 
Equation~(\ref{eq:weighted_average}) is the best estimate of the mean under the
above assumptions.

We define $\sigma$
as a square root from an unbiased estimator of the population variance 
[the Bessel correction, i.e. $(N-1)$ instead of $N$ in the denominator of
equation~\ref{eq:sigma}]
as this is the definition often adopted
in statistical software, notably in the GNU Scientific Library\footnote{\url{https://www.gnu.org/software/gsl/}\\\url{https://en.wikipedia.org/wiki/Bessel's_correction}}. 
For the purpose of variable star
search, the use of Bessel's correction has minimal practical consequences.

Standard deviation is relatively sensitive to outlier points. In many cases,
lightcurve filtering (Section~\ref{sec:filtering}) might be needed before $\sigma$ 
can serve as an efficient variable star selection tool. 
In the following paragraphs we describe ways of characterizing
lightcurve scatter that are less sensitive to outliers.

\subsection{Median absolute deviation (MAD)}
\label{sec:MAD}

The median absolute
deviation\footnote{\url{http://en.wikipedia.org/wiki/Median_absolute_deviation}},
MAD (\citealt{rousseeuw-qn-1993}, \citealt{2011ApJ...733...10R}), is a measure of scatter of observations $m_i$ defined as
\begin{equation}
{\rm MAD} = {\rm median}(|m_i-{\rm median}(m_i)|).
\end{equation}
For a Gaussian distribution
\begin{equation}
\sigma = {\rm MAD} / \Phi^{-1}(3/4) \simeq 1.4826 \times {\rm MAD}
\end{equation}
where $\Phi^{-1}(x)$ is the inverse of the cumulative distribution function
for the Gaussian distribution.
The MAD statistic is mostly insensitive to outliers \citep{2016PASP..128c5001Z};
its only disadvantage is that it is equally insensitive to real variations
that occur only occasionally, like rare eclipses of an Algol-type binary that
may show virtually constant brightness outside of the eclipses,

The use of MAD is computationally more expensive than $\sigma$ as the sorting 
needed to compute the median is a relatively slow, $\mathcal{O}(n\log{}n)$, operation
compared to calculating the average value, $\mathcal{O}(n)$. 
Here $\mathcal{O}(n\log{}n)$ ($\mathcal{O}(n)$) means that there is a
constant $C>0$ such that for any number of input measurements, $n$, the
computation will be completed in less than $C n\log{}n$ ($Cn$) steps.
It should be noted that correlation-based indices discussed below in
Sections~\ref{sec:lag1autocorr}--\ref{sec:sb} computationally depend on the order of
data points and, therefore, require the input lightcurve to be sorted in time --
an operation of $\mathcal{O}(n\log{}n)$ complexity.

\subsection{Interquartile range (IQR)}
\label{sec:IQR}

Another robust measure of scatter is the interquartile range\footnote{\url{https://en.wikipedia.org/wiki/Interquartile_range}}, 
IQR \citep{2014A&A...566A..43K}, which includes the inner 50\% of
measurement values (i.e. excludes 25\% of the brightest and 25\% of the
faintest flux measurements). To compute the IQR we first compute the median value
that divides the set of flux measurements into upper and lower halves.
The IQR is the difference between the median values computed for the upper
and lower halves of the data set. 
For the normal distribution ${\rm IQR} = 2 \Phi^{-1}(0.75) \sigma \simeq 1.349 \sigma$,
where $\Phi^{-1}(x)$ is the inverse of the cumulative distribution function
for the Gaussian distribution.
The IQR may be more appropriate than MAD~(Section~\ref{sec:MAD}) 
for measuring the width
of an asymmetric (skewed) distribution, such as the distribution of flux measurements of 
an eclipsing binary. 

\subsection{Robust median statistic (RoMS)}
\label{sec:RoMS}

The robust median statistic, RoMS, was proposed by \cite{2003AJ....126.1006E} and successfully applied for
variable star search by \cite{2007AJ....134.2067R} and \cite{2014AstBu..69..368B}.
It is defined as
\begin{equation}
{\rm RoMS} = (N-1)^{-1} \sum\limits_{i=1}^N \frac{|m_i - {\rm median}(m_i)|}{\sigma_i}.
\end{equation}
For a non-variable object,
the expected value of ${\rm RoMS}$ is around 1 as the majority of the measurements
should be within $1\sigma$ of the median value (if $\sigma$ is estimated
correctly).

\subsection{Normalized excess variance, $\sigma_{\rm NXS}^2$}
\label{sec:sigmaxs}

Normalized excess variance, $\sigma_{\rm NXS}^2$, is used in X-ray 
\citep{2009MNRAS.394.2141N,2012A&A...542A..83P,2015A&A...579A..90H,2015AJ....150...23Y}
and optical \citep{2015A&A...584A.106S} astronomy to characterize variability amplitude
in the presence of changing measurement errors. It is defined as
\begin{equation}
\sigma_{\rm NXS}^2 = \frac{1}{N \bar{m}^2} \sum\limits_{i=1}^N [ (m_i-\bar{m})^2 - \sigma_i^2 ].
\end{equation}
Here we use the symbol $\sigma_{\rm NXS}^2$ for the normalized excess
variance as this or similar symbols are widely used in the literature.
Note that $\sigma_{\rm NXS}^2$ may be negative if
the estimated errors $\sigma_i$ are larger than the actual scatter of
measurements,~$m_i$.
The fractional root mean square variability amplitude, $F_{\rm var}$, another commonly used
X-ray variability measure, is simply a square root of the
normalized excess variance: $F_{\rm var} =  \sqrt{ \sigma_{\rm NXS} }$ \citep{2003MNRAS.345.1271V}
if $\sigma_{\rm NXS}^2$ is positive. 

\cite{1993ApJ...414L..85L} note that in the presence of red noise, the
expected value of $\sigma_{\rm NXS}^2$ depends on the length of a time series. 
The value of $\sigma_{\rm NXS}^2$ estimated from a lightcurve is related to
the integral of the power spectral density (PSD) in the frequency range probed by
the observations, however this relation is complex \citep{2013ApJ...771....9A}
and depends on the PSD slope and sampling (window function).

\subsection{Peak-to-peak variability, $v$}
\label{sec:peaktopeak}

The peak-to-peak variation, $v$, can be characterized as
\begin{equation}
 v = \frac{(m_{i}-\sigma_{i})_\mathrm{max}-(m_{i}+\sigma_{i})_\mathrm{min}}{(m_{i}-\sigma_{i})_\mathrm{max}+(m_{i}+\sigma_{i})_\mathrm{min}}\,
\end{equation}
where $(m_{i}-\sigma_{i})_\mathrm{max}$ and $(m_{i}+\sigma_{i})_\mathrm{min}$
are the maximum and minimum values of the expressions $m_{i}-\sigma_{i}$
and $m_{i}+\sigma_{i}$ over the entire lightcurve.
This variability index, with minor variations in its definition,
is widely used in the radio astronomy community 
\citep[e.g.][]{1992ApJ...399...16A,2004A&A...419..485C,2008A&A...485...51H,2011IAUS..275..164F,2012AstBu..67..318M,2012ARep...56..345G}.
It is of interest
to compare $v$ with variability characteristics traditionally used in optical
and other bands. Here we use the definition of $v$ adopted by
\cite{2009AN....330..199S} and \cite{2014A&A...572A..59M}. The value of $v$ may be
negative if the measurement errors, $\sigma_{i}$, are overestimated 
(c.f.~$\sigma_{\rm NXS}^2$, Section~\ref{sec:sigmaxs})

The peak-to-peak variation may be a sensitive variability indicator if 
we believe that a lightcurve is free from outliers (thanks to high data quality or successful
filtering).
While $v$ can be computed for a lightcurve consisting of as few as two
observations, the expected value of $v$ for a non-variable source depends
strongly on the number of measurements. Monte~Carlo simulation is a practical way to
estimate expected values of $v$ for a non-variable object given a number 
of observations and their accuracy.

\subsection{Lag-1 autocorrelation, $l_1$}
\label{sec:lag1autocorr}

Photometric observations are often planned so that the time span between 
consecutive flux measurements is smaller than the variability time-scale
expected for the objects of interest. The simplest way to characterize 
the similarity of
consecutive flux measurements is to compute the first-order
autocorrelation coefficient (also known as `serial correlation
coefficient' or `lag-1 autocorrelation') of a lightcurve
\citep[e.g.][]{2011ASPC..442..447K,2011ApJ...735...68K}:
\begin{equation}
l_1 = {\sum\limits_{i = 1}^{N-1} (m_{i} - \bar{m}) (m_{i+1} - \bar{m})
       \over
       \sum\limits_{i = 1}^{N} (m_{i} - \bar{m})^2}
\end{equation}
It has been shown that, assuming that $m_{i}$ are independent measurements subject to
identically distributed measurement errors,
$l_1$ follows an asymptotically normal distribution
with the expected value of $= -1/N$ and the variance of $\simeq 1/N$, 
allowing one to assess if the obtained value of $l_1$ is consistent 
with the expected one under the above assumptions.

This simple method loses efficiency if a lightcurve is unevenly sampled
since pairs of data points widely separated in time and weakly
correlated or uncorrelated entirely contribute to the value of $l_1$ equally
with the pairs of measurements taken close in time that may be well correlated.

\subsection{Welch-Stetson variability index $I$}
\label{sec:welch}

\cite{1993AJ....105.1813W} propose a variability index, $I$, characterizing
the degree of correlation between $n$ quasi-simultaneous pairs of measurements obtained in
two filters $b$ and $v$:
\begin{equation}
I = \sqrt{ \frac{1}{n(n-1)} }\sum\limits_{i = 1}^{n} \left(\frac{b_i - \bar{b}}{\sigma_{b_i}}\right) \left(\frac{v_i - \bar{v}}{\sigma_{v_i}}\right) 
\end{equation}
where $b_i$ ($v_i$) are the measured magnitudes, $\sigma_{b_i}$ ($\sigma_{v_i}$) are the estimated errors and
$\bar{b}$ ($\bar{v}$) is the mean magnitude in filter $b$ ($v$).

Relying on the above assumption that a lightcurve contains pairs of measurements
obtained close in time (compared to the expected variability time-scale)
one can apply $I$ to a single-band lightcurve 
by dividing it into two subsamples that would mimic measurements in two
filters. One obvious way to accomplish this is to sort the lightcurve in
time, number measurements ($1, 2, 3\dots$) and assign measurements having 
odd numbers to $v$ subsample and even numbers to $b$ subsample.
In this case, $\bar{b} = \bar{v}$ may be taken to
be the mean of all $N = 2n$ observations, rather than the means of
two different samples each of size $n$.

If a single-filter lightcurve does not entirely consist of pairs of
closely-spaced points, one would like to avoid forming pairs from
measurements taken far apart in time (c.f.~$l_1$ in Section~\ref{sec:lag1autocorr}).
In that case, an additional parameter, $\Delta T_{\rm max}$, 
defines the maximum time difference between
two observations that are considered to be taken sufficiently close in time
for forming a pair. The performance of the algorithm on a given unevenly sampled
data set depends strongly on the choice of $\Delta T_{\rm max}$.
If $\Delta T_{\rm max}$ is too small, only few lightcurve points will form a
pair and contribute to $I$ rendering the index unusable.
An optimal value of $\Delta T_{\rm max}$ would be large enough to form many
measurement pairs in an unevenly sampled lightcurve but small enough to
remain sensitive to a wide range of variability time-scales as $I$ is
sensitive to variations on time-scales from $\Delta T_{\rm max}$ to the
overall duration of the lightcurve.
A histogram of the interval between observations may be useful in
selecting an appropriate $\Delta T_{\rm max}$ value for a given data set \citep{2016A&A...586A..36F}.
In our tests we use $\Delta T_{\rm max} = 2\,d$ for all the test data sets.
Isolated data points that cannot be paired with others (for a given choice of $\Delta T_{\rm max}$)
are omitted from the $I$ computation. 

Fig.~\ref{fig:simlc} and Table~\ref{tab:simlc} show how an
unevenly-sampled single-band lightcurve can be divided into subsamples to
calculate $I$ or $J$. A point is assigned to subsample b, v or counted as
`isolated' depending on the value of $\Delta T_{\rm max}$ and the order in which 
one considers the lightcurve: from the first point to the last one
(direction indicated by the top arrow in Fig.~\ref{fig:simlc}, 
the corresponding samples are named as lower case b and v) or 
the reverse direction (bottom arrow, capital B and V). 
Depending on the order, one may compute the `forward' and
`reverse' values of an index that might differ from each other because the
points are divided into pairs (assigned to b and v subsamples) in a
different way (as illustrated by Fig.~\ref{fig:simlc}). 
In our implementation of the index, the `forward' and `reverse' values
are averaged to have a single value describing a lightcurve. 
In the case of $I$ (but not $J$),
the `forward' and `reverse' values are equal if one allows a point to
be counted in multiple pairs (enter two subsamples simultaneously).

The $I$ and $J$ indices are designed to detect variability on time-scales
much longer than the typical time difference between observations forming
pairs. 
If, however, the variability time-scale is comparable to the sampling rate of
observations, the measurements in pairs may appear anticorrelated 
(correlation coefficient $l_1 \sim -1$, Section~\ref{sec:lag1autocorr})
rather than uncorrelated ($l_1 \sim 0$), resulting in near-zero or negative value of $I$ ($J$) and
rendering the index insensitive to the variations.
The actual value of detectable variability time-scale is
determined by the data and will be very different for data sets including
observations taken minutes apart and data sets that include only observations
taken on different nights.

\subsection{Stetson's $J$, $K$ and $L$ variability indices}
\label{sec:stetson}

\begin{figure}
 \centering
 \includegraphics[width=0.48\textwidth]{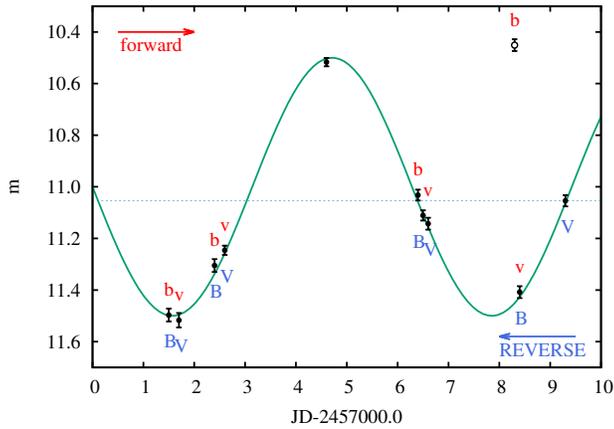}
 \caption{Single-band lightcurve simulated as $m = 11.0 + 0.5\sin({\rm JD-2457000.0}) + {\rm noise}$
 is divided into subsamples to calculate Stetson's variability
 indices (Sections~\ref{sec:welch} and \ref{sec:stetson}). 
 The arrows indicate the
 order in which lightcurve points are considered: first to last (subsample
 names in lower case) or reverse (subsample names in upper case). 
 $\Delta T_{\rm max} =1\,d$. Dashed line is the weighted average.
The weights are iteratively scaled
by the factor $f$, see equation~(\ref{eq:weightscale}), Section~\ref{sec:stetson}.
 The open circle is an `outlier' 1\,mag brighter than it should be to
 follow the sine curve.}
 \label{fig:simlc}
\end{figure}

A more robust variability index proposed by
\cite{1996PASP..108..851S} is:
\begin{equation}
J = \frac{ \sum\limits_{k = 1}^n w_k \sgn(P_k) \sqrt{|P_k|} }{ \sum\limits_{k = 1}^n w_k }
\end{equation}
where $\sgn$ is the sign function.
Here the photometric data set is divided into $n$ groups each consisting of two
or more quasi-simultaneous observations (in one or more filters) or 
a single, isolated measurement. 
A single-filter lightcurve can be divided into subsamples to mimic multi-band
data in the same way as for the $I$ index (Section~\ref{sec:welch}),
with the difference that isolated points can be kept in the analysis.
Each group consisting of one or more points is assigned a weight $w_k$.
$P_k$ is defined as
\begin{equation}
P_k = \left\{
\begin{array}{ll}
\left( \sqrt{ \frac{n_v}{n_v-1} } \frac{v_i - \bar{v}}{\sigma_{v_i}}\right) \left( \sqrt{ \frac{n_b}{n_b-1} } \frac{b_i - \bar{b}}{\sigma_{b_i}}\right) & \mbox{pair} \\
 \frac{n_v}{n_v-1} \left(\frac{v_i - \bar{v}}{\sigma_{v_i}}\right)^2 -1 & \mbox{single observation} \\
\end{array}
\right.
\end{equation}
The definition of $P_k$ can be generalized for groups containing more than
two measurements by multiplying $P_k$ (for a pair) by an additional factor of 
$\left( \sqrt{ \frac{n}{n-1} } \frac{r_i - \bar{r}}{\sigma_{r_i}}\right)$,
where $r_i$ are the observations in the third filter or subsample.
For simplicity, in the implementation of the Stetson indices used throughout this paper,
we do not consider groups containing more than two points and do not
allow a point to be counted as part of more than one group (see
Fig.~\ref{fig:simlc} and Table~\ref{tab:simlc}). 

Instead of using the weighted arithmetic mean to derive $\bar{v}$, \cite{1996PASP..108..851S}
suggests to use an iterative re-weighting procedure to down-weight potential
outlier points. After computing $\bar{v}$ as the weighted mean at the first
step, weights of all points are multiplied by a factor 
\begin{equation}
\label{eq:weightscale}
f=\left( 1 + \left(\frac{ \left| \sqrt{ \frac{n_v}{n_v-1} } \frac{v_i - \bar{v}}{\sigma_{v_i}} \right|}{a}\right)^{b}  \right)^{-1}
\end{equation}
and $\bar{v}$ is re-computed with these new weights. 
The procedure is repeated until it converges.

Many types of variable stars show continuous brightness variations
(with notable exceptions, the Algol-type eclipsing binaries and flare stars,
which remain at about constant brightness most of the time only occasionally
showing large variations).
\cite{1996PASP..108..851S} suggests to supplement $J$, which is a measure of
the degree of correlation between consecutive brightness measurements, with a
robust measure of the kurtosis (`peakedness') of the magnitude histogram:
\begin{equation}
K = \frac{1/N \sum\limits_{i=1}^N \left| \sqrt{ \frac{n_v}{n_v-1} } \frac{v_i - \bar{v}}{\sigma_{v_i}} \right|}
{\sqrt{1/N \sum\limits_{i=1}^N \left(\sqrt{ \frac{n_v}{n_v-1} } \frac{v_i - \bar{v}}{\sigma_{v_i}} \right)^2}}.
\end{equation}
For a Gaussian magnitude distribution, 
$K\xrightarrow[N \to \infty]{}\sqrt{2/{\mathrm \pi}}$ or will be less if
there is an outlier point in the lightcurve making the overall magnitude
distribution more `peaked'.

The two indices $J$ and $K$ can be combined to the index $L$ \citep{1996PASP..108..851S}:
\begin{equation}
L = \sqrt{\pi/2} J K (\sum w / w_{\rm all})
\end{equation}
where $(\sum w / w_{\rm all})$ is the ratio of the weights of all of the
lightcurve points to a total weight that the star would have if it had been
successfully measured on all images. This ratio is designed to reduce the
combined variability index $L$ value for stars with a small number of
measurements for the reasons outlined in Section~\ref{sec:filtering}.
The combined index is designed to maximize chances of detection for
well-measured continuously variable stars. It is less effective for objects
that show brightness variations only occasionally (Algol-type binaries, flare stars,
transient events).

\begin{table}
 \caption{Simulated lightcurve divided into subsamples}
 \label{tab:simlc}
 \begin{tabular}{cccc}
 \hline
 JD (days) & $m_i$ (mag) & $\sigma_i$ (mag) & subsamples \\
 \hline
 2457001.50000 & 11.497 & 0.025 & bB \\
 2457001.70000 & 11.517 & 0.028 & vV \\
 2457002.40000 & 11.305 & 0.025 & bB \\
 2457002.60000 & 11.246 & 0.018 & vV \\
 2457004.60000 & 10.517 & 0.016 &  \\
 2457006.40000 & 11.032 & 0.021 & b \\
 2457006.50000 & 11.111 & 0.020 & vB \\
 2457006.60000 & 11.143 & 0.023 & V \\
 2457008.30000 & 10.451 & 0.023 & b \\
 2457008.40000 & 11.408 & 0.023 & vB \\
 2457009.30000 & 11.054 & 0.022 & V \\
 \hline
 \end{tabular}
\end{table}

\subsection{Stetson's variability indices with time-based weighting: $J({\rm time})$, $L({\rm time})$}
\label{sec:stetsonTime}

\cite{2003ChJAA...3..151Z} and \cite{2012AJ....143..140F} suggested to weigh
the pairs used to compute Stetson's $J$ index (Section~\ref{sec:stetson}) according to
the time difference between the observations used to form a pair:
\begin{equation}
  w_i = \exp {\left( -\frac{t_{i+1}-t_i}{\Delta t} \right) },
\end{equation}
where $t_i$ is the time of observation $i$ and $\Delta t$ is the median of
all pair time spans $(t_{i+1}-t_i)$.
This weighting scheme eliminates the need to choose a specific maximum
allowed time difference ($\Delta T_{\rm max}$, Section~\ref{sec:welch}) for forming a pair.

\subsection{Stetson's variability indices with a limit on the magnitude difference in a pair: $J({\rm clip})$, $L({\rm clip})$}
\label{sec:stetsonClip}

The example presented in Fig.~\ref{fig:simlc} shows that it is undesirable 
to form a pair that would include an outlier point. Considering the assumption that a
lightcurve contains pairs of observations taken close in time (compared to
the expected variability time-scale) one can discard 
from the calculation of $I$ (Section~\ref{sec:welch}) or $J$ index (Section~\ref{sec:stetson}) pairs with magnitude
difference between the points greater than a few times the measurement uncertainty.
In our tests, we do not form pairs from measurements that differ by
more than five times their combined uncertainty, no matter how close in time
the two measurements are taken.

\subsection{Consecutive same-sign deviations from the mean magnitude (CSSD)}
\label{sec:cssd}

\cite{2000AcA....50..421W} and \cite{2009MNRAS.400.1897S} suggested to use the number of
groups, CSSD,
containing three consecutive measurements that are brighter or fainter than
the mean (or median) magnitude by at least a factor of $c \sigma$ as a variability
indicator. Typically, the value of $c$ is set to 2 or 3.
In the algorithm implementation tested in this work, we choose $c = 3$, 
replace $\sigma$ with the MAD value scaled to $\sigma$ (Section~\ref{sec:MAD})
and use the median as the baseline magnitude level, in order to make the index more robust against
outliers.  Following \cite{2009MNRAS.400.1897S} we normalize the number of
groups by $(N-2)$, where $N$ is the number of points in a lightcurve.

\subsection{Excursions, $E_x$}
\label{sec:excursions}

\cite{2008ApJS..175..191P} and \cite{2014ApJS..211....3P} point out that
ground-based photometric time series can often be naturally divided into groups
(scans) -- dense series of observations separated by long gaps. If the variability
time-scale is longer than the duration of an individual scan, 
average 
magnitudes will differ from scan to scan.
Combining observations within a scan to form a single estimate of brightness has an
obvious advantage of giving a more accurate estimate (compared to an individual
measurement) at the expense of degraded time resolution.

To compare mean magnitudes of the scans, one could perform the
analysis of variance \cite[ANOVA; e.g.][]{KenneyKeeping}. However, 
a lightcurve obtained with a ground-based telescope is likely to violate 
the assumptions behind the parametric form of this test.
The variance of measurements may differ between the scans (if the observations
combined in different scans were performed at different elevations or weather 
conditions). The distribution of measurements may be non-Gaussian due to outliers.
It is tempting to use a non-parametric test (like Mood's median test)
to compare scans without having a pre-conception about the measurement error
distribution. 
However, when applied to a typical ground-based photometric
data set, such a test would give the (mathematically correct) answer that the
majority of stars are variable due to night-to-night photometric zero-point
variations. 

In the algorithm implementation tested here, we use the absolute difference
between the median magnitudes of scans normalized by their combined MADs (Section~\ref{sec:MAD})
and averaged over all pairs of scans in a lightcurve to form the variability index $E_x$.
In practice, the exact way a lightcurve is split into scans has a
strong impact on the usefulness of this variability test for a given data set. 
We divide the lightcurve in scans according
to a predefined maximum time difference. This implies that each scan may
have a different number of points. For each scan we compute the median and
MAD scaled to $\sigma$ (Section~\ref{sec:MAD}) of the observed magnitudes during this scan. 
The index $E_x$ is computed according to the equation:
\begin{equation}
E_x=\frac{2}{N_{\rm scan}(N_{\rm scan}-1)}\sum_{i=1}^{N_{\rm scan}-1}\sum_{\substack{j>i,}}^{N_{\rm scan}}
\frac{\left|{\rm median}_i-{\rm median}_j\right|}{\sqrt{\sigma_i^2+\sigma_j^2}}
\end{equation}
where $N_{\rm scan}$ is the number of scans, 
$N_{\rm scan}(N_{\rm scan}-1)/2 = C^2_{N_{\rm scan}}$ is the number
of two-scan combinations in the data set,
${\rm median}_i$ and $\sigma_i$
corresponds to median and MAD scaled to $\sigma$ of the $i$th scan and the
same notation is used for the $j$th scan, respectively.

\subsection{The von~Neumann ratio $\eta$}
\label{sec:vonneumann}

The ratio of the mean square successive difference to the distribution
variance was discussed by \cite{vonneumann1941,vonneumann1942} as an indicator of
independence for a series of observations. It is defined as:
\begin{equation}
\eta = \frac{\delta^{2}}{\sigma^{2}} =  
\frac{\sum\limits_{i=1}^{N-1}(m_{i+1} - m_{i})^2 / (N - 1)}{ \sum\limits_{i=1}^N (m_i-\bar{m})^2  / (N - 1)}.
\end{equation}
It remains useful even if the observations are drawn from a non-Gaussian
distribution as long as it is nearly-symmetric \citep{Lemeshko,Strunov}.

The ratio $\eta$ quantifies the smoothness of a time series.
\cite{2009MNRAS.400.1897S} employed $\eta$ as a variability indicator,
noting that since photometric time series measurements 
do not follow a Gaussian distribution, in practice, the cut-off value for
selecting variable objects cannot be determined a~priori (as in the case of
$\chi^2$, Section~\ref{sec:chi2}). 
One may use $1/\eta$ as a variability indicator to have larger
values of the index corresponding to a greater likelihood of an object being
variable as is the case with the other variability indices discussed here.

\subsection{Excess Abbe value $\mathcal{E}_\mathcal{A}$}
\label{sec:EA}

\cite{2014AnA...568A..78M} discussed the Abbe value $\mathcal{A} = \eta/2$ and the
excess Abbe value
\begin{equation}
\mathcal{E}_\mathcal{A} \equiv \overline{ \mathcal{A}_\mathrm{sub} } - \mathcal{A}
\end{equation}
where $\overline{ \mathcal{A}_\mathrm{sub} }$ is the mean of
$\mathcal{A}_\mathrm{sub~i}$
values computed for all measurements $m_i$ obtained at times $t_i$.
Each $\mathcal{A}_\mathrm{sub~i}$ is computed over the sub-interval 
 $[t_i-\frac{1}{2}\Delta T_\mathrm{sub}, t_i+\frac{1}{2}\Delta T_\mathrm{sub}]$
($\Delta T_\mathrm{sub} < \Delta T$, the overall duration of time series).
The choice of $\Delta T_\mathrm{sub}$ determines the minimum time-scale of
variability that may be detected by comparing 
$\mathcal{A}_\mathrm{sub~i}$ to $\mathcal{A}$. 
$\mathcal{E}_\mathcal{A}$ may be useful to identify unusual behaviour in
well-sampled lightcurves. A large number of measurements 
($>5$ in our implementation) should be taken within the time interval $\Delta T_\mathrm{sub}$
from each point to accurately determine $\mathcal{A}_\mathrm{sub~i}$.

\subsection{$S_B$ variability detection statistic}
\label{sec:sb}

The $\chi^2$ statistic applied to photometric time series data considers
only the distribution of the measured magnitudes ignoring the information 
on when these measurements were obtained. Thus the $\chi^2$ statistic cannot distinguish between
the cases where small-scale deviations in one direction from the mean value are
randomly distributed across the lightcurve from the cases where many of the
same-sign deviations are concentrated
around a specific time (the second case is less likely to occur by chance).

\cite{2013AnA...556A..20F} suggested a variability detection statistic that 
combines the advantages of scatter-based and correlation-based variability
indices. It is based on the `alarm' statistic used by
\cite{2006MNRAS.367.1521T} to assess the quality of fitting binary
lightcurve models to observational data. \cite{2012MNRAS.420.1333A} applied a
similar statistic to detect the Blazhko effect in lightcurves of RR~Lyrae stars.
The variability detection statistic is defined as
\begin{equation}
S_B=\left(\frac{1}{NM}\right)\sum_{i=1}^{M}\left(\frac{r_{i,1}}{\sigma_{i,1}}
+\frac{r_{i,2}}{\sigma_{i,2}}+\dots+\frac{r_{i,k_{i}}}{\sigma_{i,k_{i}}}
\right)^2
\end{equation}
where $N$ represents the total number of data points in the lightcurve and $M$ is the number of groups of
consecutive residuals of the same sign from a constant-brightness light
curve model, $r_{i,j} = |m_i - \bar{m}|$ ($j$~is the running number in the
group containing $k_{i}$ same-sign deviations from the mean, $\bar{m}$) and
$\sigma_{i,j}$ are the uncertainties corresponding to $r_{i,j}$.

\section{Test data sets}
\label{sec:testdata}

To compare the relative power of the indices (Section~\ref{sec:indices}) 
in identifying variable objects we use seven
photometric data sets containing a large number of known variable stars
(Table~\ref{tab:testdatasummary}).
The data sets represent a range of sampling patterns and measurement
accuracies. 
Due to the diversity of instruments and reduction strategies, 
the data sets are characterized by a variety of numbers of badly measured
objects that contaminate the lists of candidate variables.
Overall, the selected data sets
should represent a range of observing conditions typically found in
ground-based variability surveys.

The data sets used for our variability indices test were
previously searched for variability and contain 1097 known variable objects.
While preparing this publication we manually checked the lightcurves of all
stars standing out in any of the variability indices
plotted against the mean magnitude (Fig.~\ref{fig:idxsample}). 
We were able to identify 124 variable stars that
were missed in the original searches.
New variable stars\footnote{Information about known
variable stars was extracted from the AAVSO International Variable Star Index
(VSX; \citealt{2006SASS...25...47W}; \url{https://www.aavso.org/vsx}) and VizieR service
(\url{http://vizier.u-strasbg.fr/}). Variable stars were considered `new'
if no information about their variability could be found in these services.}
were found in each one of the test data sets. 
This highlights the fact that variability detection techniques used in previous searches can be improved 
by adding (a combination of) the variability indices considered here (Section~\ref{sec:indices}).

\begin{table}
   \caption{Test data sets. $N_{\rm var}$~--~number of variable stars identified in the
data set, $N_{\rm stars}$~--~total number of stars and
$N$~--~maximum number of lightcurve points.}
   \label{tab:testdatasummary}
   \begin{tabular}{c r@{~~}c@{~~}c@{~~}l c}
   \hline\hline
Dataset    & $N_{\rm var}$/$N_{\rm stars}$ & $N$ & Time range  & $m_{\rm lim}$ & Sec. \\
   \hline
TF1       & 271/21543 & 3900 & 2012-05-14 to 2013-08-19 & 18\,$R$          & \ref{sec:kpsobs} \\
TF2       &  51/ 8438 & 8004 & 2014-09-05 to 2014-11-22 & 16\,$R$          & \ref{sec:kpsobs} \\
Kr        & 235/29298 & 1171 & 2012-08-13 to 2012-10-18 & 17\,$V^\dagger$  & \ref{sec:krasnoyarsk} \\
W1        &  80/ 2615 &  242 & 2006-06-14 to 2006-07-24 & 19\,$I$          & \ref{sec:westerlund1} \\
And\,1    & 124/29043 &  132 & 2011-10-31 to 2013-05-23 & 14\,$V^\dagger$  & \ref{sec:nmw} \\
SC20      & 465/30265 &  268 & 1997-10-05 to 2000-11-24 & 21\,$I$          & \ref{sec:ogle} \\
66\,Oph   &  26/ 6337 &  227 & 1976-02-04 to 1995-08-19 & 17\,$B^\ddagger$ & \ref{sec:66oph} \\
\hline
   \end{tabular}
\vspace{-8pt}
\begin{flushleft}$^\dagger$~Unfiltered magnitude calibrated against $V$ zero-point.\\
$^\ddagger$~Photographic magnitudes calibrated against $B$ zero-point.\end{flushleft}
\end{table}

\subsection{The Kourovka Planet Search (TF1, TF2)}
\label{sec:kpsobs}

As our test data we used observations of two dense sky fields in the
Galactic plane conducted within the framework of the Kourovka Planet Search
\citep{2016MNRAS.461.3854B}.
The first field (TF1) was observed with 
the MASTER-II-Ural telescope at the Kourovka Astronomical
Observatory of the Ural Federal University ($\varphi=57\degr$~N,
$\lambda=59\degr$~E). The mean full width at half-maximum (FWHM) seeing at the site is $3$\,arcsec.
The telescope consists of a pair of Hamilton catadioptric
tubes (400\,mm $f$/2.5) on a single equatorial mount Astelco NTM-500 without autoguiding. 
Each tube is equipped with $4098\times4098$ pixels
Apogee Alta~U16M CCD giving an image scale of 1.85\,arcsec~pixel$^{-1}$ in a
$2\times2$~deg$^2$ field. 
The field TF1 is centred at
$\upalpha_{\text{J2000}}$=20:30:00 $\updelta_{\text{J2000}}$=+50:30:00 (Cygnus). The
main observing set of TF1 was completed during short and bright nights from
2012~May to August. We obtained 3900 frames with an exposure time of 50\,s in the $R$ filter.
The time interval between consequent frames was about 1.5~min.
TF1 was observed for 90\,h in the $R$ band (36 nights) 
with an average duration of 2.5\,h per observing run.

The second field (TF2) was observed with the Rowe-Ackermann Schmidt astrograph (RASA) telescope 
(279\,mm $f$/2.2) 
at the Acton Sky Portal private observatory ($\varphi=43\degr$~N,
$\lambda=71\degr$~W). The telescope is equipped with a $3352\times2532$ pixels SBIG
STF-8300M CCD
which provides an image scale of 1.79\,arcsec~pixel$^{-1}$ in a
$1.2\times1.6$~deg$^2$ field. 
The typical seeing at the site is $2$\,arcsec.
TF2 is centred at 
$\upalpha_{\text{J2000}}$=02:47:00 $\updelta_{\text{J2000}}$=+63:00:00
(Cassiopeia). 
The RASA telescope obtained about 8000 frames of TF2 in 2014~September--November  
during all available clear nights. Observations were performed in the $R$ filter with 50\,s exposure
time. The time interval between the consequent frames is 1~min. The field
was observed for 130\,h (18 nights) with an average duration of 7.2\,h per night. 

Before processing the data, we had to filter out some of the images because
not all of them were obtained in optimal weather conditions. 
We use the standard deviation of image pixel counts $\sigma_{\text{pix}}$ as
an indicator of weather conditions. 
The value of $\sigma_{\text{pix}}$ varies smoothly from image to
image in photometric nights. In the presence of clouds $\sigma_{\text{pix}}$
value of a particular image noticeably decreases (or increases if the clouds are
lit by the moonlight). 

We used the console version of the {\scshape astrometry.net}
application~\citep{Lang2010} to set the correct World Coordinate System
parameters in the FITS header of each image. {\scshape IRAF} package
\citep{Tody1986} is then used to perform 
dark frame subtraction and division by the flat-field.
Dark frames are taken before each observing night. Flat-field images are
taken during the dawn. The {\scshape phot/apphot} task is used to
perform aperture photometry in each frame with 
aperture size and sky background level adjusted for each image. 
The aperture radius is set to $0.8\,{\rm FWHM}$ 
of the stellar PSF in the frame.
A total of 21500 and 8500 stars were measured in TF1 and TF2 fields, respectively.
The {\scshape astrokit} software \citep{2014AstBu..69..368B} is used to correct 
for the star brightness variations caused by changing atmospheric
transparency. The program selects for each star an individual
ensemble of reference stars having similar magnitudes and positions in the frame. 
We achieved photometric accuracy of 
0.005--0.05\,mag in the interval 11--16\,mag for data from
the MASTER-II-Ural telescope. 
For the RASA telescope data, we 
achieved precision of 0.006--0.08\,mag in the magnitude interval 11--16\,mag for the TF2 field.    
These lightcurve data were originally searched for
variability by \cite{2015PZP....15....7P}.

\subsection{Krasnoyarsk SibSAU 400\,mm telescope (Kr)}
\label{sec:krasnoyarsk}

A $2.3\times2.3$~deg$^2$ field centred at
$\upalpha_{\text{J2000}}$=22:50:00
$\updelta_{\text{J2000}}$=+52:00:00 (Lacerta)
was observed with the 400\,mm $f$/2.3 telescope of the Siberian State Aerospace
University using the $3056\times3056$~pixels (2.7\,arcsec~pixel$^{-1}$) unfiltered CCD camera
FLI~ML09000.
The telescope is installed on top of the University building in the city of
Krasnoyarsk. The turbulence caused by the building results in
$7$--$8$\,arcsec seeing. The observing site is affected by light
pollution. A total of 1171 30\,s exposures of the field were obtained in
2012~August--October. After applying bias, dark and flat-field corrections
using the {\scshape MaxIm~DL} software the images were loaded into 
{\scshape VaST}\footnote{\url{http://scan.sai.msu.ru/vast/}}
\citep{2005ysc..conf...79S} for photometric analysis.

After comparing results of aperture and PSF-fitting photometry 
performed using {\scshape SExtractor} \citep{1996A&AS..117..393B} with {\scshape PSFEx}
\citep{2011ASPC..442..435B}, we discovered that for the brightest stars in
the field, the aperture photometry is about a factor of 2 more accurate
than PSF photometry probably 
due to the insufficient accuracy of the reconstruction of 
PSF variations across the field. We applied six iterations of 
{\scshape SysRem} \citep{2005MNRAS.356.1466T,2015MNRAS.454..507M}  
to remove effects of these variations
and bring scatter of PSF-photometry lightcurves for the bright stars to the
level of scatter obtained with the aperture photometry. For the final
analysis we used {\scshape SysRem}-corrected PSF-photometry lightcurves as they provide
better measurement accuracy for the faint stars compared to fixed-aperture
photometry. Only isolated objects with {\scshape SExtractor} flag = 0 and
measured on at least 200 images were considered. The instrumental magnitude
scale is calibrated to Cousins $R = V - 1.09*(r-i) - 0.22$
\citep{2005AJ....130..873J}
computed from UCAC4/APASS $V$, $r$ and $i$
magnitudes \citep{2013AJ....145...44Z,2016yCat.2336....0H} of 2644 stars in the field.
These images were originally investigated by
\cite{2013PZP....13...12L} and \cite{2016PZP....16....4L}
who used {\scshape VaST} with {\scshape SExtractor} in the aperture
photometry mode and identified variable objects using the $\sigma$--mag plot.

\subsection{LCO 1\,m Swope telescope (W1)}
\label{sec:westerlund1}

Observations of the Galactic super star cluster Westerlund\,1
were obtained during 17 nights between 2006 June~14 and July~24 using the 1\,m
$f$/7 Henrietta Swope
telescope at Las Campanas Observatory, Chile by \cite{2007AJ....133.2696B}
who identified 129 new variable stars in the field using image subtraction.
A $1200\times1200$ pixels section of the $2048\times3150$ SITe CCD
(0.435\,arcsec~pixel$^{-1}$) corresponding to $8.7$\,arcmin field of view was 
read to increase cadence. The initial image processing steps including
overscan-correction, linearity-correction and flat-fielding were performed in
{\scshape IRAF}.
We re-processed 242 $I$-band images 
(including some rejected from the original study due to poor seeing)
with {\scshape VaST},
performing PSF-fitting photometry using {\scshape SExtractor} and {\scshape PSFEx}. 
The magnitude scale was calibrated using $I$-band magnitudes of 1276 stars in the
field measured by \cite{2007AJ....133.2696B}.
We considered only isolated objects 
({\scshape SExtractor} flag $=0$) detected on $\ge100$ images to minimize
the effects of crowding. 
Three cycles of {\scshape SysRem} are applied to the data.
From the list of \cite{2007AJ....133.2696B} we
select 78 objects showing detectable variability in the $I$-band and pass our
selection criteria. We add two previously unknown variable objects found during our tests
(Table~\ref{tab:newvar}, Fig.~\ref{fig:newvarlightcurves}).

\begin{table*}
 \caption{New variable stars found in the test data (the full table is available online)}
 \label{tab:newvar}
 \begin{tabular}{c l c c c c c}
 \hline
 Name & Alias & $\upalpha_{\text{J2000}}$ $\updelta_{\text{J2000}}$ & Mag. range & Type & Period ($d$) & Epoch \\
 \hline
ogle\_17707  &  LMC\_SC20\_17707           &  05:44:59.79 $-$70:53:45.1  & 17.85-17.95\,I &  SR     &  86  &  max 2451164.798  \\
ogle\_33977  &  LMC\_SC20\_33977           &  05:45:07.10 $-$70:38:57.3  &  19.00-19.3\,I &  E:/CEP:  &  8.220  &  min 2450842.770  \\
ogle\_14141  &  LMC\_SC20\_14141           &  05:45:07.95 $-$70:56:55.9  & 17.65-17.75\,I &  SR     &  34.5  &  max 2450726.854  \\
 \end{tabular}
\end{table*}

\begin{figure*}
 \centering
 \includegraphics[width=0.9\textwidth,clip=true,trim=0cm 22.6cm 0cm 0cm]{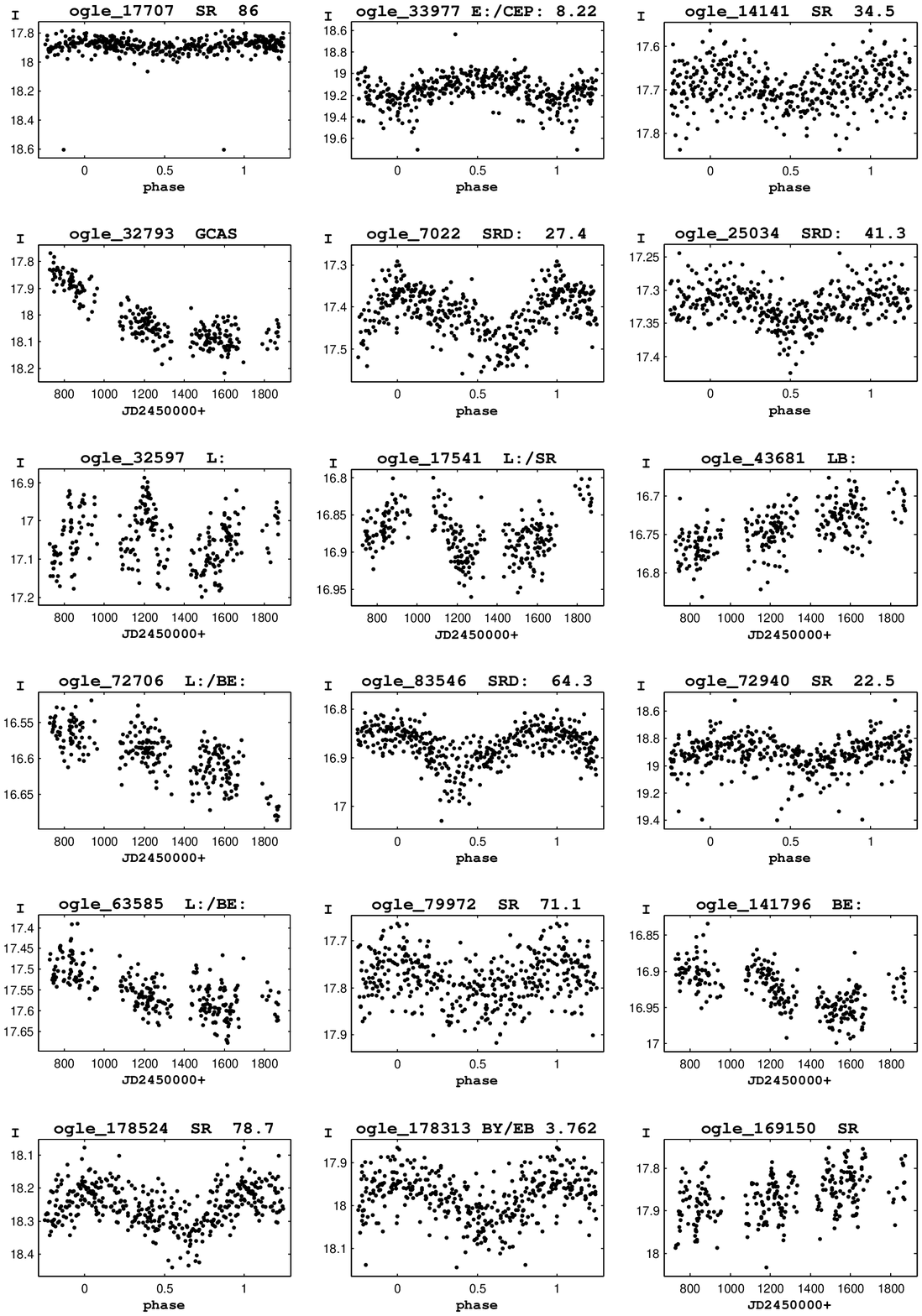}
 \caption{Lightcurves of new variable stars found in the test data sets (the complete figure is available online). 
The magnitudes measured in a band indicated in the top-left corner of each
panel are plotted as a function of time (Julian day) for irregular or phase
for periodic variables. The title of each panel indicate the object
identifier in Table~\ref{tab:newvar}, its variability type and period in days (if applicable).}
 \label{fig:newvarlightcurves}
\end{figure*}

\subsection{The New Milky Way survey (And\,1)}
\label{sec:nmw}

The New Milky Way survey\footnote{\url{http://scan.sai.msu.ru/nmw/}}
\citep{2014ASPC..490..395S}
hunted for bright ($V<13.5$\,mag) transients near
the Galactic plane using a Canon\,EF $f=135$\,mm ($f$/2) telephoto lens attached to an
unfiltered $3352\times2532$ SBIG ST-8300M CCD camera
(8.4\,arcsec/pix, $8\times6$~deg$^2$ field). 
The observations were conducted in 2011--2013.
We used 132 images of the field centred at 
$\upalpha_{\text{J2000}}$=23:00:00
$\updelta_{\text{J2000}}$=+50:00:00
(And\,1)\footnote{The And\,1 field fully includes 
the deeper SibSAU 400\,mm Lacerta field described in
Section~\ref{sec:krasnoyarsk}.}
reprocessed with {\scshape VaST} and {\scshape SExtractor} in the
aperture photometry mode accepting blended stars for the analysis 
({\scshape SExtractor} flag $\le3$).
Since the CCD chip is blue-sensitive, APASS $V$-band magnitudes of
~1200 UCAC4 
stars within the field of view
are used for magnitude calibration. Three cycles of {\scshape SysRem}
were applied to the data in order to mitigate 
systematic effects caused by chromatic aberration of the lens and 
changing atmospheric extinction across the large field of view.
We used three {\scshape SysRem} cycles as adding more cycles did not further 
improve (reduce) lightcurve scatter for the majority of objects in this
data set.

Lightcurves and images of all objects that stand out in index versus mag plots
were visually inspected for variability. We identified 91 known and 33
previously unknown variable stars (Table~\ref{tab:newvar}, Fig.~\ref{fig:newvarlightcurves}). 
The list of detectable variable stars in the field may be incomplete as we
accepted only those red objects showing slow irregular variability that are
either matched with a known variable star or their variability can be
confirmed from ROTSE-I/NSVS \citep{2004AJ....127.2436W} or
SuperWASP \citep{2010A&A...520L..10B} archival data. This should safeguard
us from mistaking for real variability any residual colour-related systematics 
not removed by {\scshape SysRem}.
An example of such residual systematic variation is the dip around
JD2456000 visible in the lightcurves of many red (SR and LB type) variables
in the field (Fig.~\ref{fig:newvarlightcurves}).
As the final check we repeat the processing using elliptical aperture of
size and orientation that are individually tuned for each object 
({\scshape SExtractor} parameter {\scshape MAG\_AUTO}). This allows us to
recover the flux of defocused red stars at the cost of reducing photometric
accuracy for the well-focused point sources and make sure that for the
selected variable star candidates the {\scshape MAG\_AUTO} lightcurve shape
is not contradicting the one obtained with a circular aperture of a size
fixed for all objects in a given image.

\subsection{OGLE-II (LMC\_SC20)}
\label{sec:ogle}

The Optical Gravitational Lensing Experiment 
utilizes the dedicated 1.3\,m Warsaw telescope at the Las Campanas
Observatory, Chile, to conduct a photometric survey of dense stellar fields in the Magellanic Clouds
and Galactic bulge \citep{1997AcA....47..319U}.
We extract data from the second phase of the experiment OGLE-II
PSF $I$-band photometry database \citep{2005AcA....55...43S}. 
For the variability index tests we
select one field in the Large Magellanic Cloud, LMC\_SC20, which is least
affected by crowding. To keep the number of selected sources below the limit
of 50000 imposed by the database's web-interface and retrieve only
high-quality lightcurves, we selected sources having the percentage of
good measurements ${\rm Pgood} \ge 98$. In total, 30265 sources in this field
satisfy the selection criteria each having from 262 to 268 photometric measurements.
%
The data set contains 168 variable stars 20 of which (see Table~\ref{tab:newvar}, Fig.~\ref{fig:newvarlightcurves}) 
were not previously known. The new variable stars were identified by visual
inspection of the lightcurves standing out in variability index versus mag
plots. To make sure the detected variability is not caused by nearby
bright variable stars, we visually checked PSF-fitting lightcurves of stars located within
$20$\,arcsec of each of the new variables. Only stars brighter than
the variable were considered and no limit on the percentage of good
measurements was applied.

The use of a fixed centroid position when conducting photometry may
introduce spurious long-term variability if the measured star has a detectable proper
motion. If the DIA is used, the moving star will
have a characteristic dipole shape in the residual image, resulting in
detection of two spurious variable sources apparently changing brightness 
in opposite directions \citep{2001MNRAS.327..601E}. To make sure the
variability of ogle\_43681, ogle\_63585 and ogle\_72706 
is not caused by the proper motion, we {(i)}\,check that there are no
records in the OGLE-II DIA catalogue by \cite{2001AcA....51..317Z} within
$3$\,arcsec of the new variables and {(ii)}\,manually check OGLE-II DIA
lightcurves of nearby sources to make sure none of them show brightness
trends mirroring the new variables.

\subsection{Digitized photographic plates (66\,Oph)}
\label{sec:66oph}

Photographic images of the sky obtained in late 19th and 20th centuries
contain a wealth of information about historical 
positions \citep[e.g.][]{2010AJ....140.1062L,2010A&A...509A..62V,2014A&A...572A.104R,2013SoSyR..47..203B} and brightness 
\citep[e.g.][]{2008AcA....58..279K,2013PASP..125..857T,2014ARep...58..319S} of celestial
objects. The efficient use of this information requires it to be converted
to a digital form using purpose-built digitizing machines
\citep{2006SPIE.6312E..17S,2012ASPC..461..315D}
or a commercially-available flatbed scanners capable of working with
transparent materials \citep{2007A&A...471.1077V,2009ASPC..410..111S,2014aspl.conf..127T}.

We use an Epson Expression 11000XL flatbed scanner
operating at 2400\,dpi resolution 
($1.4$\,arcsec/pix, 16\,bits per pixel colour depth)
to digitize a $1.26 \times 1.26$~deg$^2$ area 
centred at 
$\upalpha_{\text{J2000}}$=17:57:44.7
$\updelta_{\text{J2000}}$=+04:59:54
(66\,Oph field; \citealt{2010ARep...54.1000K})
on 227 photographic plates obtained in 1976--1995 with the 40\,cm astrograph.
The digitized images were processed with {\scshape VaST} following the
procedure described by \cite{2014aspl.conf...79S}.
APASS $B$-band photometry of ~1600 UCAC4 
stars in the magnitude range $B$=10--16 is used to calibrate the instrumental magnitude
scale using the relation between aperture photographic and photoelectric
magnitudes proposed by \cite{2005MNRAS.362..542B}.
We identify 23 variable stars including five not previously known
(Table~\ref{tab:newvar}, Fig.~\ref{fig:newvarlightcurves}) by means of period search and visual inspection of
lightcurves standing out in the magnitude versus $\sigma_w$ plot.

\subsection{Lightcurve filtering}
\label{sec:filtering}

Often raw photometric data have to be pre-processed before 
computing the variability indices discussed in Section~\ref{sec:indices}.
This may include {(i)}\,removing outliers from a lightcurve (possibly by applying
iterative $\sigma$-clipping or median filtering); {(ii)}\,removing systematic effects from a set of
lightcurves by applying local zero-point corrections
\citep[e.g.][]{2014MNRAS.442.2381N} and/or the {\scshape SysRem} algorithm, 
decorrelating each lightcurve with external parameters such as
airmass, seeing, object position on a CCD, detector temperature
\citep[e.g.][]{2009PhDT.......230P,2010ApJ...710.1724B,2010ApJ...716L..36L,2011AJ....141..166H,2012ApJS..201...36B,2015sf2a.conf..277G,2016A&A...588A..56B}
or de-trending the lightcurves if one is
interested only in fast variability \cite[e.g.][]{2005MNRAS.356..557K,2015AN....336..125W}.

A smaller-than-expected number of detections is an indirect indication of many photometry
problems including the object being close to
an image edge, a cosmetic defect, a bright star, a detection or saturation limit.
Objects systematically affected by any of these problems can be removed from
the analysis by discarding lightcurves having less than a given number of
points. The obvious disadvantage is that together with problematic objects,
one may discard a transient object that appears only on a
small number of images. The power of discarding lightcurves with a small
number of measurements to improve the overall quality of a photometric data set 
might be the reason why `variable star detection' and
`optical transient detection' are traditionally considered as two
separate technical problems.

From all the data sets considered in this work we discard lightcurves having 
fewer than 40 points, unless indicated otherwise.
We apply no $\sigma$-clipping to the test data, however we note that 
it can considerably improve performance of variability indices
that are not robust to outliers.
The {\scshape SysRem} algorithm is applied to the data sets described in
Sections~\ref{sec:krasnoyarsk}, \ref{sec:westerlund1} and \ref{sec:nmw}.
For the other data sets it does not lead to a noticeable
reduction in lightcurve scatter.

\subsection{Simulated data sets}
\label{sec:simulated}

The data sets described above (Section~\ref{tab:testdatasummary}) include in total
1251 variable stars of various types, but this list still provides us limited
coverage of a possible range of variability amplitudes and time-scales.  
We overcome this limitation by adding simulated variability to the test data.
Following \cite{2012A&A...548A..48E}, we use lightcurves of non-variable
stars as realistic photometric noise models. 
This approach has an advantage over simple 
bootstrapping\footnote{Here by bootstrapping we mean shuffling JD--magnitude pairs in a lightcurve 
to eliminate any correlated variability. The methods is often used to assess
the significance of a periodogram peak \citep[e.g.][]{2011MNRAS.413.2696B}.} in that it preserves the
correlated nature of the noise. It naturally requires a set of constant
stars to have multiple realizations of the noise process while the bootstrapping
can be applied to an individual lightcurve.

From each set of lightcurves described above we remove the known variable stars
and introduce artificial variability to the remaining stars that are presumed 
to be constant. Among these constant stars there are both well-measured ones
and some affected by blending or other sources of large photometric errors.
According to the simulation parameters, each star has a 1\% chance
to be variable with a random peak-to-peak amplitude
uniformly distributed between 0 and 1.0\,mag.
The simulation is done in two versions: 
in version 1 all variables are assumed to be periodic while in the second
version they are all assumed to be aperiodic.

We model periodic variability by adding a simple sine signal
\citep[e.g.][]{2016MNRAS.455..626M} to the observed lightcurve of a constant star.  
The signal phase is chosen randomly for each simulated variable star.
The frequency of the sine signal is drawn from a uniform
random distribution in the range 0.05 -- 20~d$^{-1}$. This results in a large
fraction of variables with periods $<1$~d approximately resembling 
the period distribution typically found in the Galactic field.

To simulate non-periodic variability we sum-up 10000 sine waves with
logarithmically spaced frequencies 
in the range 0.0001--1000\,d$^{-1}$ and
having random phases. 
The amplitude of each sine wave is the square root of
the power spectrum value. 
If the real and imaginary parts of the Fourier transform of the lightcurve
are independent and vary according to the Gaussian distribution
\citep{2012PhDT.......389C}, the resulting power of the sine waves 
is varying according to 
the $\chi_2^2$ distribution with 2 degrees of freedom 
\citep{1995A&A...300..707T,2013MNRAS.433..907E}
around the expected values. The expected values in our simulation are 
defined by a power law with the slope of $-1$ 
\citep[e.g.][]{2014MNRAS.445..437M}.
The exact choice of the power law slope in the range $-0.5$ to $-3.0$ 
has minimal effect on the following discussion.
The simulations are repeated 1000 times for each data set and the averaged
results are reported.

\section{Comparison technique}
\label{sec:comparison_tech}

To select the variability index that is the most efficient
in identifying variable stars, we compute the indices defined in
Section~\ref{sec:indices} for all lightcurves in the test data sets
(Section~\ref{sec:testdata}).
The variable objects have to be distinguished from two broad types of
interlopers:
non-variable objects and objects with corrupted photometry.
To quantify the performance of each index following 
\cite{2011ApJ...735...68K} and \cite{2014MNRAS.439..703G}, we compute
the completeness $C$ and purity $P$:
\begin{equation}
C = \frac{\mathrm{Number \,\, of \,\, selected \,\, variables}}{\mathrm{Total \,\, number \,\, of \,\, confirmed \,\, variables}}
\end{equation}
\begin{equation}
P = \frac{\mathrm{Number \,\, of \,\, selected \,\, variables}}{\mathrm{Total \,\, number \,\, of \,\, selected \,\, candidates}}
\end{equation}
as well as the fidelity
$F_1$-score\footnote{The three parameters are often referred to as 
`recall' or `sensitivity' or `true positive rate', `precision' and `$F$-factor'
for $C$, $P$ and $F$, respectively. See
\url{https://en.wikipedia.org/wiki/Precision_and_recall}
\url{https://en.wikipedia.org/wiki/F1_score}
\url{https://en.wikipedia.org/wiki/Information_retrieval}} 
which is the harmonic mean of the two parameters:
\begin{equation}
F_1 = 2 (C \times P) / (C + P).
\end{equation}
$F_1$ reaches a maximum of 1.0 for a perfect selection when all confirmed
variables and no false candidates pass the selection criteria while $F_1=0$ 
if no confirmed variables are selected. 

For each variability index, $A$, described in Section~\ref{sec:indices} we compute
its expected value $\bar{A}$ and dispersion $\sigma_A$ as functions of magnitude.
The operation is performed for each data set that includes real
(Section~\ref{sec:testdata}) and simulated (Section~\ref{sec:simulated}) variable
objects.
For each point in the magnitude versus index (mag--$A$) plot we use points within 
$\pm0.25$\,mag from it to compute $\bar{A}$ as a
median of indices within the magnitude bin. 
If the bin contains $<40$ points, its width is increased to include at least 40 points. 
The expected dispersion, $\sigma_A$, is computed as the MAD scaled to $\sigma$ (Section~\ref{sec:MAD})
for the points in the bin. 
After completing these computations for all the points in the
magnitude--index plot, the estimated values of $\bar{A}$ and 
$\sigma_A$ are smoothed with a simple running-average.
The robust estimators of $\bar{A}$ and $\sigma_A$ are necessary 
considering that a bin is likely to contain variable or badly measured
objects that have variability index values not typical 
for constant stars. 

\begin{figure}
 \centering
 \includegraphics[width=0.47\textwidth]{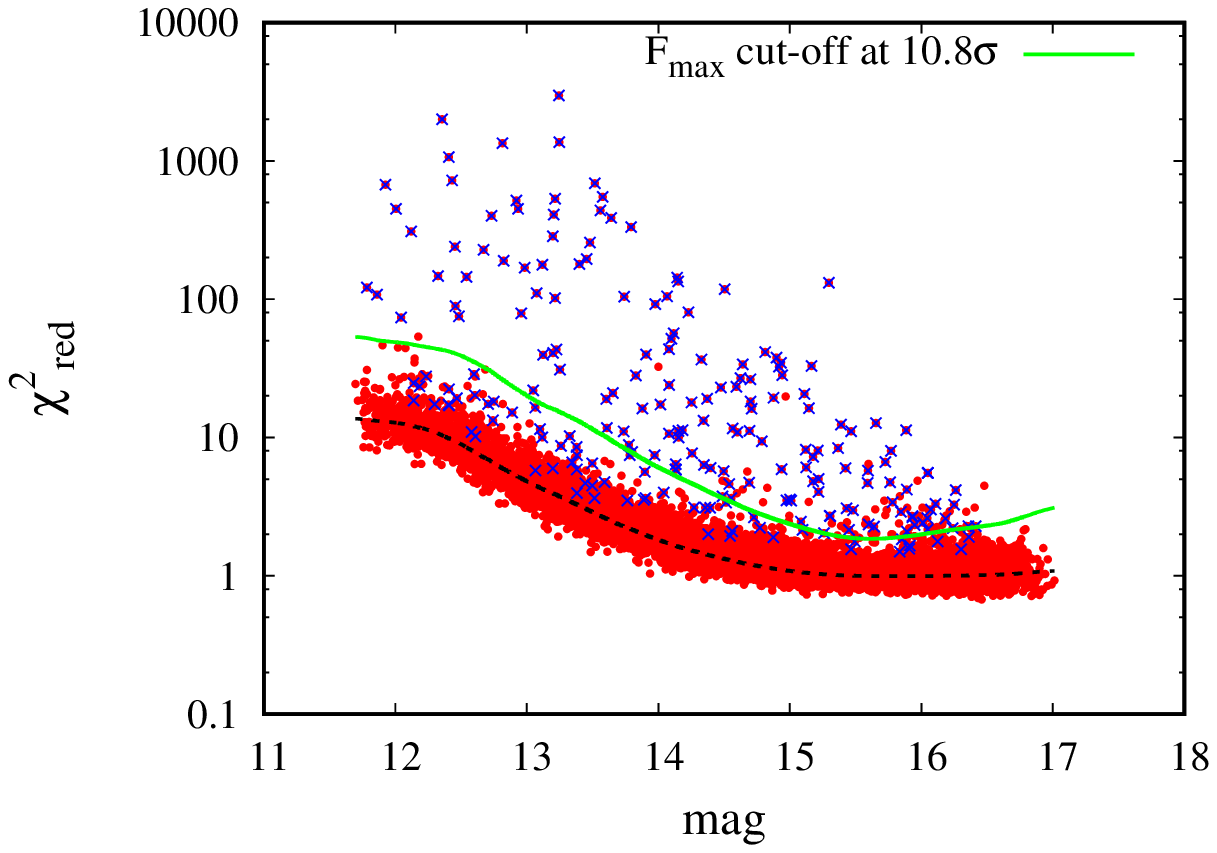}
 \includegraphics[width=0.47\textwidth]{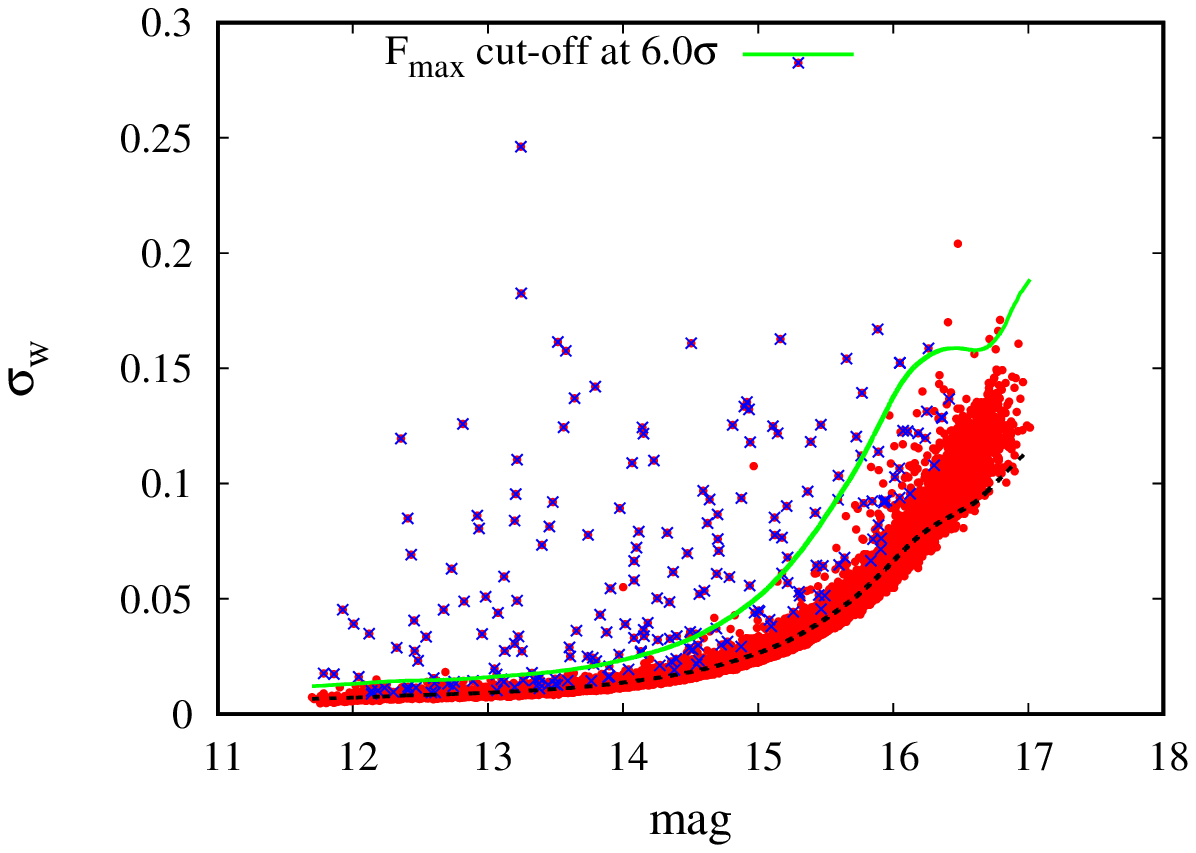}
 \includegraphics[width=0.47\textwidth]{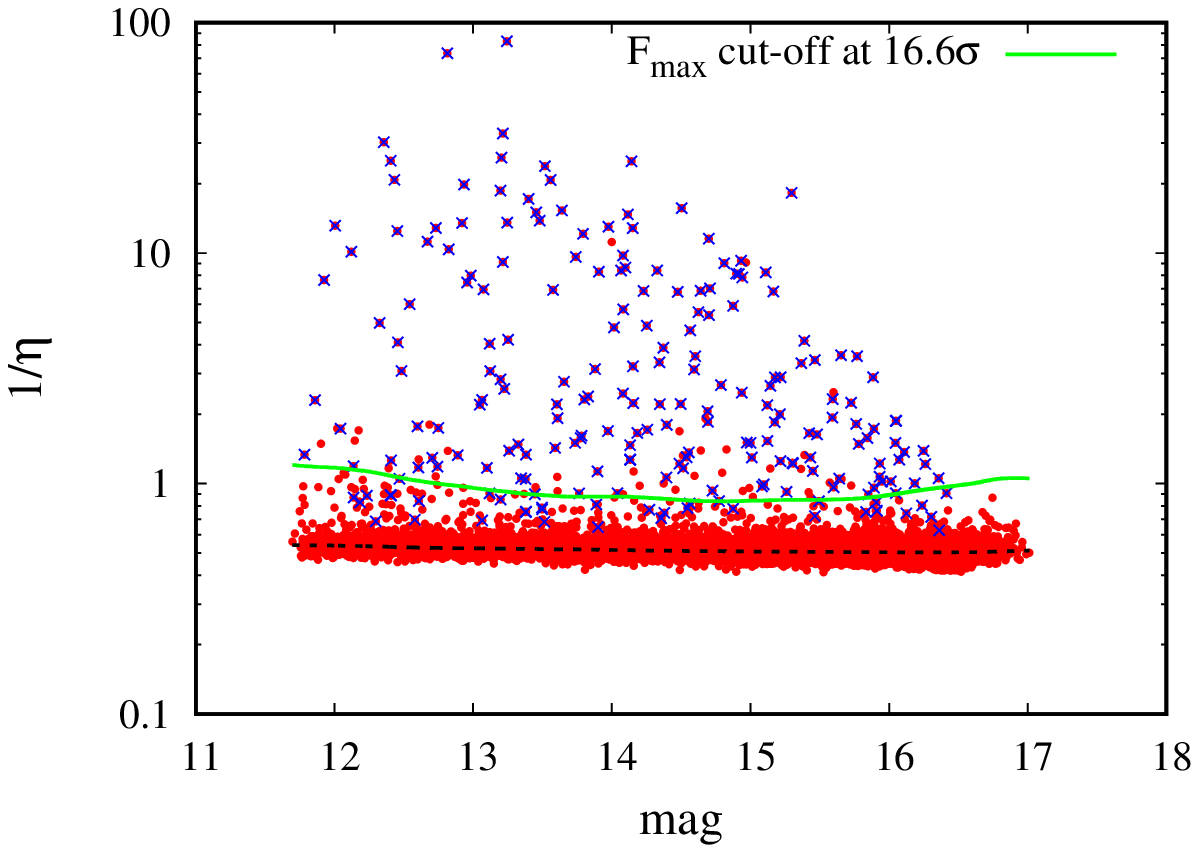}

 \caption{Variability indices $\chi_{\rm red}^2$ (Section~\ref{sec:chi2}),
$\sigma_w$ (Section~\ref{sec:sigma}) and $1/\eta$ (Section~\ref{sec:vonneumann})
plotted as a function of magnitude for the Krasnoyarsk data set
(Section~\ref{sec:krasnoyarsk}). 
Variable stars are marked with 'x'.
The curves represent the expected value
of $\chi_{\rm red}^2$, $\sigma_w$ and $1/\eta$ for a given magnitude and the selection threshold
corresponding to the best trade-off between the completeness and purity of the 
candidates list ($F_{\rm max}$; see Section~\ref{sec:comparison_tech},
Fig.~\ref{fig:CPFsample}). The index--magnitude plots for the other indices
and data sets may be found online at \url{http://scan.sai.msu.ru/kirx/var_idx_paper/}}
 \label{fig:idxsample}
\end{figure}

Variable star candidates are selected as objects having a variability index
value deviating by more than $a\sigma_A$ from the value $\bar{A}$ expected
at this magnitude, where $a$ is a factor defining the variability detection
threshold (Fig.~\ref{fig:idxsample}).
This approach is similar to the one employed by \cite{2011MNRAS.413.2696B} who
selected periodic variable stars using a cut in false alarm probability (FAP)-period space. The authors used the median and MAD 
as robust estimators of the expected FAP value and its scatter as a function of a period.
Unlike \cite{2010ApJ...723..737V}, we compare the variability indices
not at some specific cut-off level $a$ common for all indices, but instead
choose the optimal value of $a$ individually for each index as described below.

\begin{figure*}
 \centering
 \includegraphics[width=0.33\textwidth]{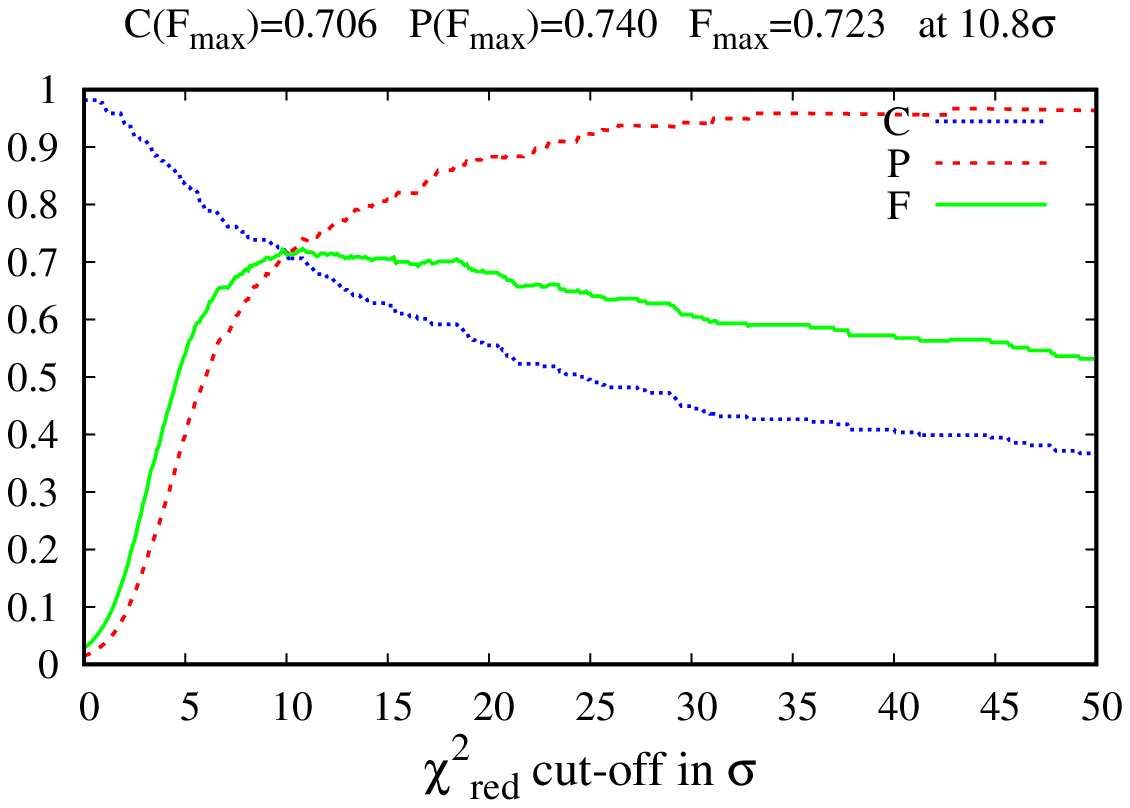}
 \includegraphics[width=0.33\textwidth]{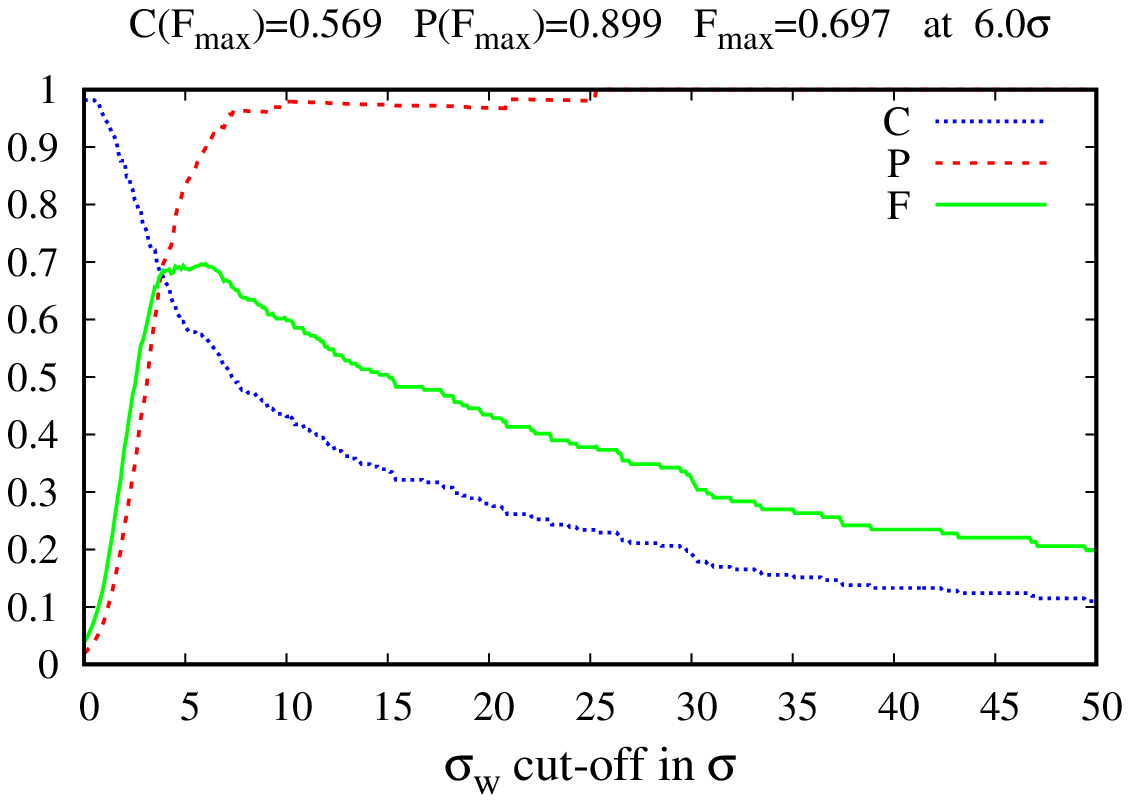}
 \includegraphics[width=0.33\textwidth]{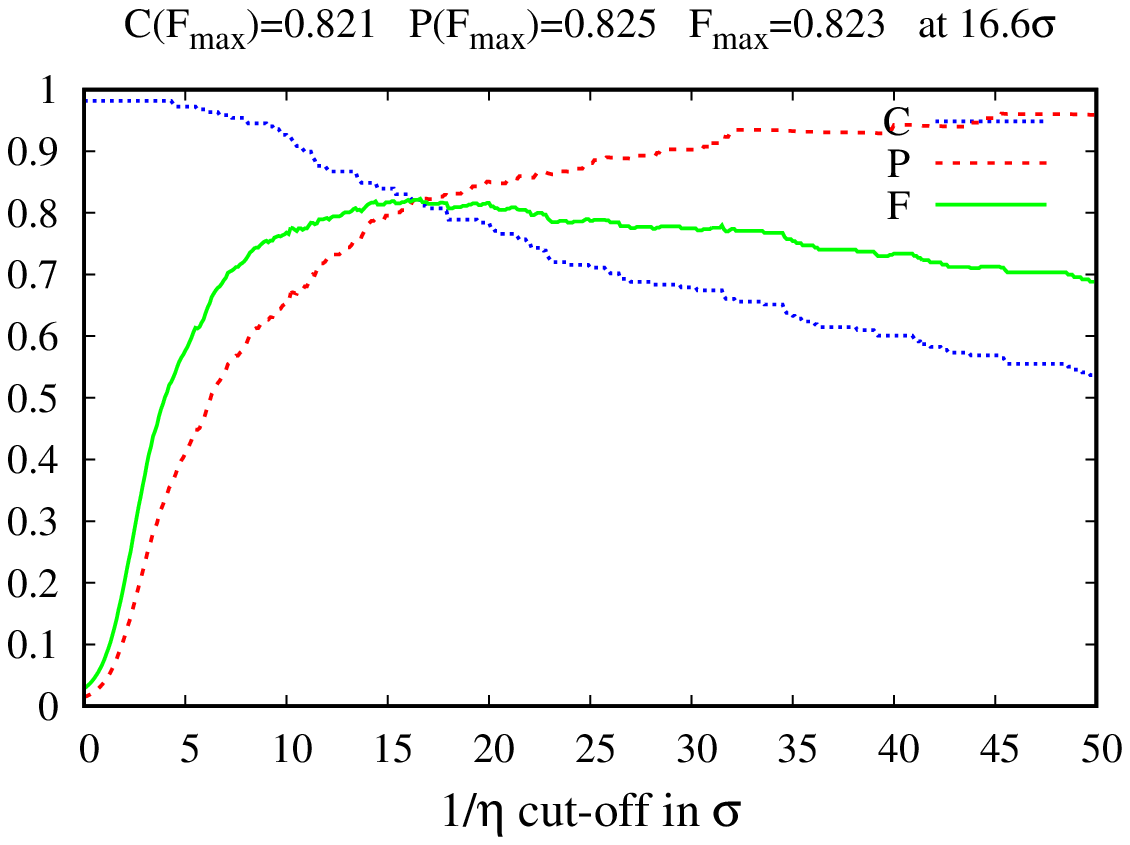}

 \caption{Variable star selection completeness (C),
purity (P) and $F_1$-score (F; see Section~\ref{sec:comparison_tech}) as a function of selection threshold for the
variability indices $\chi_{\rm red}^2$ (Section~\ref{sec:chi2}),   
$\sigma_w$ (Section~\ref{sec:sigma}) and $1/\eta$ (Section~\ref{sec:vonneumann})
computed for the Krasnoyarsk data set (Section~\ref{sec:krasnoyarsk}). C-, P-,
and $F_1$-score plots for the other indices and data sets may be found online
at \url{http://scan.sai.msu.ru/kirx/var_idx_paper/}}
 \label{fig:CPFsample}
\end{figure*}

For each index and data set we compute $C$, $P$ and $F_1$ parameters as functions of $a$ (Fig.~\ref{fig:CPFsample}).
For some optimal value of $a$, $F_1$ reaches the maximum, $F_{1~\rm max}$,
corresponding to a trade-off between the completeness and purity of the selected
list of candidates.
We consider the index with the highest value of $F_{1~\rm max}$ as the most efficient in
selecting true variable stars in a given data set.
By comparing results for various data sets
(Sections~\ref{sec:testdata} and \ref{sec:simulated}), we draw general conclusions
about which indices perform better under a wide range of observing
conditions (Section~\ref{sec:resdisc}).
Since $F_1$ characterizes only the list of selected candidates and does not
consider the rejected ones, we also report a fraction of objects that 
do not pass the selection (at the cut-off value
corresponding to $F_{1~\rm max}$), $R$, as a supplementary measure of variability
index performance.


\begin{table*}
   \caption{Performance of variability indices in selecting real variable stars}
   \label{tab:FRreal}
   \begin{tabular}{c c@{~~}c c@{~~}c c@{~~}c c@{~~}c c@{~~}c c@{~~}c c@{~~}c cc}
   \hline\hline
                       & \multicolumn{2}{c}{TF1} & \multicolumn{2}{c}{TF2}  & \multicolumn{2}{c}{Kr}           & \multicolumn{2}{c}{Westerlund\,1} & \multicolumn{2}{c}{And\,1} & \multicolumn{2}{c}{LMC\_SC20} & \multicolumn{2}{c}{66\,Oph} &        & \\
Index                  & $F_{1~\rm max}$ & $R$ & $F_{1~\rm max}$ & $R$ & $F_{1~\rm max}$ & $R$ & $F_{1~\rm max}$ & $R$ & $F_{1~\rm max}$ & $R$    & $F_{1~\rm max}$ & $R$  & $F_{1~\rm max}$ & $R$     & Sec. & Ref. \\
   \hline
\multicolumn{15}{c}{Scatter-based indices} \\
 $\chi_{\rm red}^2$     & 0.110 & 0.902   & 0.076 & 0.884   & 0.723 & 0.993   & 0.270 & 0.969   & 0.556 & 0.996   & 0.284 & 0.996   & 0.192 & 0.992  & \ref{sec:chi2}            & (a) \\
 $\sigma_w$             & 0.114 & 0.899   & 0.076 & 0.879   & 0.697 & 0.995   & 0.264 & 0.950   & 0.544 & 0.996   & 0.254 & 0.994   & 0.155 & 0.988    & \ref{sec:sigma}           & (b) \\
 ${\rm MAD}$            & 0.161 & 0.927   & 0.086 & 0.940   & 0.710 & 0.994   & 0.287 & 0.940   & 0.582 & 0.996   & 0.483 & 0.994   & 0.375 & 0.996    & \ref{sec:MAD}             & (c) \\
 ${\rm IQR}$            & 0.162 & 0.927   & 0.086 & 0.951   & 0.726 & 0.994   & 0.298 & 0.945   & 0.608 & 0.996   & 0.470 & 0.992   & 0.383 & 0.997    & \ref{sec:IQR}             & (d) \\
 ${\rm RoMS}$           & 0.130 & 0.917   & 0.070 & 0.922   & 0.729 & 0.993   & 0.270 & 0.963   & 0.563 & 0.996   & 0.382 & 0.993   & 0.381 & 0.997    & \ref{sec:RoMS}            & (e) \\
 $\sigma_{\rm NXS}^2$   & 0.026 & 0.198   & 0.012 & 0.197   & 0.047 & 0.731   & 0.059 & 0.522   & 0.032 & 0.752   & 0.034 & 0.754   & 0.324 & 0.992    & \ref{sec:sigmaxs}         & (f) \\
 $v$                    & 0.053 & 0.835   & 0.032 & 0.901   & 0.347 & 0.996   & 0.140 & 0.984   & 0.450 & 0.997   & 0.049 & 0.899   & 0.098 & 0.994    & \ref{sec:peaktopeak}      & (g) \\
\multicolumn{15}{c}{Correlation-based indices} \\
 $l_1$                  & 0.370 & 0.992   & 0.175 & 0.999   & 0.400 & 0.995   & 0.188 & 0.935   & 0.569 & 0.996   & 0.470 & 0.996   & 0.450 & 0.998    & \ref{sec:lag1autocorr}    & (h) \\
 $I$                    & 0.116 & 0.896   & 0.082 & 0.891   & 0.819 & 0.993   & 0.281 & 0.973   & 0.611 & 0.994   & 0.500 & 0.996   & 0.341 & 0.997    & \ref{sec:welch}           & (i) \\
 $J$                    & 0.144 & 0.927   & 0.079 & 0.931   & 0.819 & 0.993   & 0.286 & 0.977   & 0.628 & 0.994   & 0.448 & 0.994   & 0.368 & 0.998    & \ref{sec:stetson}         & (j) \\
 $J({\rm time})$        & 0.152 & 0.931   & 0.081 & 0.932   & 0.819 & 0.992   & 0.291 & 0.975   & 0.659 & 0.995   & 0.519 & 0.996   & 0.410 & 0.998    & \ref{sec:stetsonTime}     & (k) \\
 $J({\rm clip})$        & 0.134 & 0.922   & 0.074 & 0.917   & 0.788 & 0.993   & 0.267 & 0.977   & 0.587 & 0.995   & 0.375 & 0.991   & 0.364 & 0.997    & \ref{sec:stetsonClip}     & (d) \\
 $L$                    & 0.169 & 0.923   & 0.092 & 0.942   & 0.821 & 0.992   & 0.283 & 0.979   & 0.706 & 0.996   & 0.470 & 0.994   & 0.571 & 0.997    & \ref{sec:stetson}         & (j) \\
 ${\rm CSSD}$           & 0.231 & 0.957   & 0.105 & 0.977   & 0.014 & 0.008   & 0.034 & 0.013   & 0.008 & 0.007   & 0.011 & 0.012   & 0.008 & 0.001    & \ref{sec:cssd}            & (l) \\
 $E_x$                  & 0.181 & 0.973   & 0.090 & 0.998   & 0.347 & 0.997   & 0.159 & 0.983   & 0.500 & 0.997   & 0.357 & 0.996   & 0.263 & 0.998    & \ref{sec:excursions}      & (m) \\
 $1/\eta$               & 0.549 & 0.991   & 0.414 & 0.992   & 0.823 & 0.993   & 0.378 & 0.982   & 0.588 & 0.997   & 0.471 & 0.997   & 0.424 & 0.999    & \ref{sec:vonneumann}      & (n) \\
 $\mathcal{E}_\mathcal{A}$ & 0.154 & 0.962   & 0.156 & 0.995   & 0.434 & 0.997   & 0.250 & 0.989   & 0.151 & 0.994   & 0.228 & 0.997   & 0.133 & 0.999    & \ref{sec:EA}              & (o) \\
 $S_B$                  & 0.146 & 0.893   & 0.092 & 0.891   & 0.766 & 0.992   & 0.261 & 0.982   & 0.463 & 0.993   & 0.303 & 0.989   & 0.246 & 0.995    & \ref{sec:sb}              & (p) \\
\hline
 $\alpha_1$             & 0.112 & 0.878   & 0.078 & 0.875   & 0.782 & 0.994   & 0.245 & 0.961   & 0.639 & 0.995   & 0.441 & 0.994   & 0.426 & 0.997    & \ref{sec:pca}             & (d) \\
   \hline
   \end{tabular}
   \renewcommand{\arraystretch}{1.0}
\begin{flushleft}References: (a)~\cite{2010AJ....139.1269D},
(b)~\cite{2008AcA....58..279K},
(c)~\cite{2016PASP..128c5001Z},
(d)~this work,
(e)~\cite{2007AJ....134.2067R},
(f)~\cite{1997ApJ...476...70N},
(g)~\cite{1989ApJ...340..150B},
(h)~\cite{2011ASPC..442..447K},
(i)~\cite{1993AJ....105.1813W},
(j)~\cite{1996PASP..108..851S},
(k)~\cite{2012AJ....143..140F},
(l)~\cite{2009MNRAS.400.1897S},
(m)~\cite{2014ApJS..211....3P},
(n)~\cite{2009MNRAS.400.1897S},
(o)~\cite{2014AnA...568A..78M},
(p)~\cite{2013AnA...556A..20F}.
\end{flushleft}
\end{table*}

\section{Results and Discussion}
\label{sec:resdisc}

\subsection{Overall performance comparison}
\label{sec:overall}

Fig.~\ref{fig:idxsample} presents the variability~index--magnitude plots.
The completeness, purity and $F_1$-score as a function of the cut-off limit,
$a\sigma_A$, are presented in Fig.~\ref{fig:CPFsample}.
Table~\ref{tab:FRreal} lists the highest $F_1$-score, $F_{1~\rm max}$ and the
corresponding fraction of rejected objects, $R$, for each index and data set
described in Section~\ref{sec:testdata}.
Tables~\ref{tab:FRsimperiodic} and \ref{tab:FRsimnonperiodic} (available online)
present this information for the simulated data sets discussed in
Section~\ref{sec:simulated}.

While performance of each individual index varies
considerably between the data sets, the correlation-based indices $I$
(Section~\ref{sec:welch}), $J$, $L$
(including their time-weighted and clipped versions; Sections~\ref{sec:stetson},
\ref{sec:stetsonTime}, \ref{sec:stetsonClip}) and $1/\eta$
(Section~\ref{sec:vonneumann}) typically provide higher $F_{1~\rm max}$ values
than scatter-based indices. Among the scatter-based indices the
IQR~(Section~\ref{sec:IQR}) and MAD~(\ref{sec:MAD}) show the highest
$F_{1~\rm max}$ values with RoMS~(\ref{sec:RoMS}),
$\sigma_w$ (\ref{sec:sigma}) and 
$\chi_{\rm red}^2$ (\ref{sec:chi2}) falling slightly behind due to their sensitivity to
individual outlier measurements. The $l_1$ (\ref{sec:lag1autocorr}), 
$S_B$ (\ref{sec:sb}), $E_x$ (\ref{sec:excursions}) and $E_A$ (\ref{sec:EA}) perform well in some data sets,
but not in the others and, therefore, cannot be recommended as
general-purpose variability detection statistics. The indices $\sigma_{\rm NXS}^2$
(\ref{sec:sigmaxs}) and $v$ (\ref{sec:peaktopeak}) typically reach smaller
$F_{1~\rm max}$ values compared to the other scatter-based indices.

The CSSD index (\ref{sec:cssd}) in our implementation appears
practically useless for variable objects detection. The requirement for
three consecutive data points to be $2 \sigma_{\rm MAD}$ brighter or
fainter than the median brightness where $\sigma_{\rm MAD}$ is the $\sigma$
scaled from the lightcurve MAD (\ref{sec:MAD}) appears to be too strict.
Indeed, \cite{2000AcA....50..421W} used individual measurement errors to
compute CSSD while \cite{2009MNRAS.400.1897S} used the lightcurve
$\sigma$ to compute CSSD (similar to our implementation), 
but it was only one of the many lightcurve features used simultaneously 
for variability detection in that work.

The $1/\eta$ appears to be the best compromise
index as it performs better than most of the other discussed indices in all
tested data sets (real and simulated) judging both from $F_{1~\rm max}$ and $R$
values. The $1/\eta$ index is sensitive only to variability on time-scales
longer than the sampling time which causes it to miss fast variables in
sparsely sampled data sets like LMC\_SC20 (\ref{sec:ogle}) and 66\,Oph
(\ref{sec:66oph}). If the data set has no measurements taken very close in
time (compared to the fastest expected variability time-scale), the $1/\eta$
index sensitive to slow variations should be complemented with a
scatter-based index such as the IQR~(\ref{sec:IQR}) that would pick fast
variables missed by $1/\eta$.

\subsection{Performance based on the number of points in~a~lightcurve}

The results presented in 
Tables~\ref{tab:FRreal}--\ref{tab:FRsimnonperiodic}
allow us to identify indices that perform well in all the test data sets
(Section~\ref{sec:overall}). All the test data sets are well sampled containing
hundreds to thousands of observations. The question remains how well these
indices perform on lightcurves containing a smaller number of points?
This is especially interesting considering that the alternative period
search-based methods of variability detection (not considered in this work) 
are ineffective for lightcurves having a small number of
points. A rule of thumb is that $\gtrsim100$ points randomly sampling a
lightcurve (\citealt{2013MNRAS.434.3423G} suggest $\gtrsim200$ for CRTS sampling)
are often sufficient to determine a variable star period. 
A smaller number of points may be sufficient if the sampling is favourable or
the range of possible periods is constrained by prior knowledge of the variability type.
If the number of observations is too small to attempt a periodicity search, 
variability indices are the best hope for identifying variable stars 
among such undersampled lightcurves.

\begin{figure}
 \centering
 \includegraphics[width=0.48\textwidth]{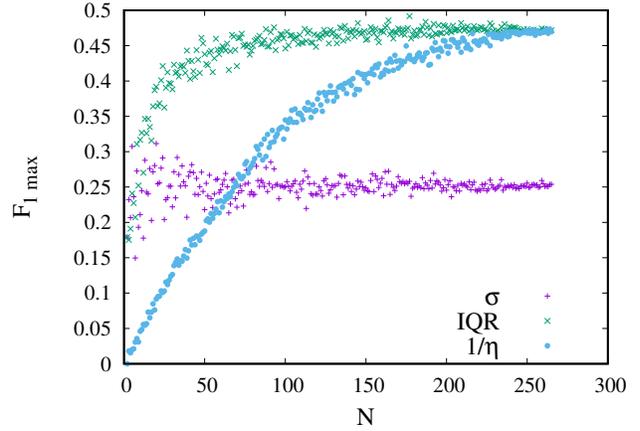}
 \caption{The $F_{1 \rm max}$-score as a function of
 the number of lightcurve points. For each $N$, the random selection of $N$ points is
 repeated 10 times and the median $F_{1 \rm max}$ value is plotted.}
 \label{fig:npoints}
\end{figure}

To test this we use the OGLE-II LMC\_SC20 data set described
in Section~\ref{sec:ogle} that is characterized by quasi-random sampling (i.e.
it includes a small number of measurements taken on the same night). We
randomly select a subset of $N$ observations from the LMC\_SC20 data to
construct an artificial data set and test how many known variables can be
recovered using the same technique as applied to the full data set
(Section~\ref{sec:comparison_tech}). The results of index comparison are presented in
Fig.~\ref{fig:npoints}. While $\sigma_w$ does not show a strong dependence on
the number of points, the $F_{1 \rm max}$-score of $1/\eta$ and $l_1$ linearly
increases with increasing number of points in a lightcurve. The IQR at
$N\lesssim15$ shows $F_{1 \rm max}$ values similar to $\sigma_w$, but it
shows larger $F_{1 \rm max}$ values for a larger number of points. 
The reason for IQR being more efficient than $\sigma_w$ for large $N$ is that the
IQR is insensitive to outlier measurements. Stetson's $J$ (and $L$) indices,
MAD and RoMS also behave similarly to the IQR as these indices can
characterize the lightcurve scatter while remaining relatively insensitive
to outliers. The Welch--Stetson $I$ index becomes useful only for a large
number of points because only in this way there are lightcurve points
obtained close enough in time to form pairs (unlike the $J$ index, $I$
cannot take into account the individual, unpaired measurements). $S_B$ does
not show a strong dependence of its $F_{1 \rm max}$ values on the number of
points, while $F_{1 \rm max}$ values of $E_A$ slowly increase with
increasing $N$. Overall, we can conclude that the indices characterizing
the lightcurve scatter perform well even on undersampled lightcurves while
the indices that are purely correlation-based linearly increase their
effectiveness with increasing number of lightcurve points.

\subsection{Correlation between variability indices}

Many of the variability indices considered above reflect the same
information, just in a slightly different way. Consider, for example, 
the three versions of Stetson's $J$ index described in
Sections~\ref{sec:stetson}--\ref{sec:stetsonClip} which
essentially differ from each other only in the relative weights assigned to
various pairs of observations.

To quantify the degree of similarity between the indices we computed the
Pearson product-moment
correlation coefficient, $r$, for all possible pairs of indices using the full
data sets (i.e. indices computed for variable and non-variable objects were
considered together). 
The linear Pearson correlation coefficient of two variables measures the
degree of linear dependence between the variables. It is defined as the
ratio of the covariance of the two variables to the product of their
standard deviations. 
It is a direct measure of how well two sample populations vary jointly. 
It ranges from $-1$ (total anticorrelation) to 1 (total correlation). 
A zero value corresponds to a lack of linear correlation 
(however, non-linear correlations may exist). 

The majority of variability indices considered in Section~\ref{sec:indices} are
strongly ($r>0.8$) correlated with each other. The exceptions are 
$l_1$, CSSD, $1/\eta$, $\mathcal{E}_\mathcal{A}$. This 
suggests that the correlated variability indices
reflect mostly the same information. This is understandable considering that
the indices quantifying the degree of correlation between consecutive
brightness measurements are also sensitive to the overall lightcurve
scatter (with the exception of $l_1$).

\subsection{Principal Component Analysis}
\label{sec:pca}

To further quantify the relative importance of the variability indices 
and to search for a possible linear combination of indices that may be a
better variability indicator than any individual index we performed the
principal component analysis \citep[PCA;][]{pearson1901}.

PCA is an unsupervised, non-parametric method
that provides a linear orthogonal transformation of a data set into a new
base, where the data
variance (assumed to represent the useful information) is highlighted. 
The new set of (uncorrelated) `optimal' axes is called the principal  
components (PCs). The original data can be expressed as a linear
combination of the PCs. Usually, very few of the PCs (even two to three of them)
are capable of describing the data in terms of variance without a
significant loss of information. This dimensionality reduction/data
compression is the reason why PCA is very effective in extracting
information from huge data sets.
However, the results should be interpreted with caution, since the data may
not reflect uncorrelated physical phenomena.
PCA is extensively used in astronomy, e.g. in applications on stellar
spectra \citep{1998MNRAS.298..361B,2007A&A...467.1373R}, on galaxy spectra
\citep{2004AJ....128..585Y,2012A&A...538A..38K}, on spectroscopic imaging
\citep{2009MNRAS.395...64S}, etc. It was suggested as a variability
detection tool for photometric data sets containing quasi-simultaneous 
multi-colour observations \citep{2006ASPC..349...15E,2012MNRAS.424.2528S}.
 
The PCA implementation on an 
$(\textit{n} \ {\rm observations}) \times (\textit{m} \ {\rm features})$ 
data set involves 
{(i)}\,the construction of either (usually) the data
variance-covariance matrix or the correlation matrix, 
{(ii)}\,the calculation of the respective eigenvectors PCi (the principal components) 
and {(iii)}\,the calculation of the admixture coefficients $\alpha_{i}$, which
are the data
coordinates on the new axes.

Thus, each original observation $x$ is decomposed on to the new set of axes PCi as

$$x = \sum_{i=1}^{m} \alpha_{i} \cdot {\rm PCi}$$

The first principal component PC1 summarizes the majority of the data
variance
(the most widespread information), PC2 summarizes the
majority of the rest of the data variance, being uncorrelated to PC1, etc.
It is expected that low-order PCs correspond to rare/weak processes, noise, etc. \citep{tso}.

PCA was applied to each of the test data sets (Section~\ref{sec:testdata}). 
The variability indices of the sample's stars were
normalized by their expected value $\bar{A}$ and scatter $\sigma_A$ as a function of magnitude as
discussed in Section~\ref{sec:comparison_tech}.
Since the indices represent different, albeit
often correlated, characteristics and PCA is data-dependent, 
we performed a zero mean and unit variance 
standardization
prior to the analysis. Additionally, the
variance-covariance matrix of the data was used. PCA was implemented in IDL
(PCOMP procedure).

\begin{figure}
 \centering
 \includegraphics[width=0.45\textwidth]{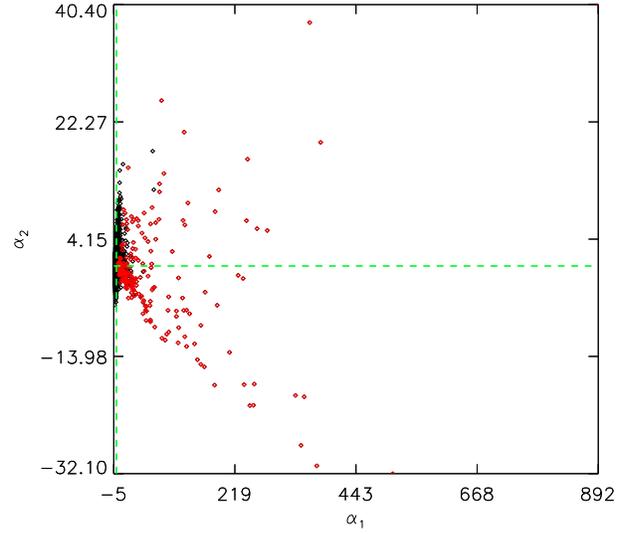}
 \caption{The admixture coefficients corresponding to PC1 ($\alpha_{1}$),
PC2 ($\alpha_{2}$) for the Kr data set (Section~\ref{sec:krasnoyarsk}). 
 Variable stars are marked in red. Similar plots for PC3 ($\alpha_{3}$) 
and the other data sets may be found at \url{http://scan.sai.msu.ru/kirx/var_idx_paper/}}
 \label{fig:admixture}
\end{figure}

\begin{figure}
 \centering
 \includegraphics[width=0.48\textwidth]{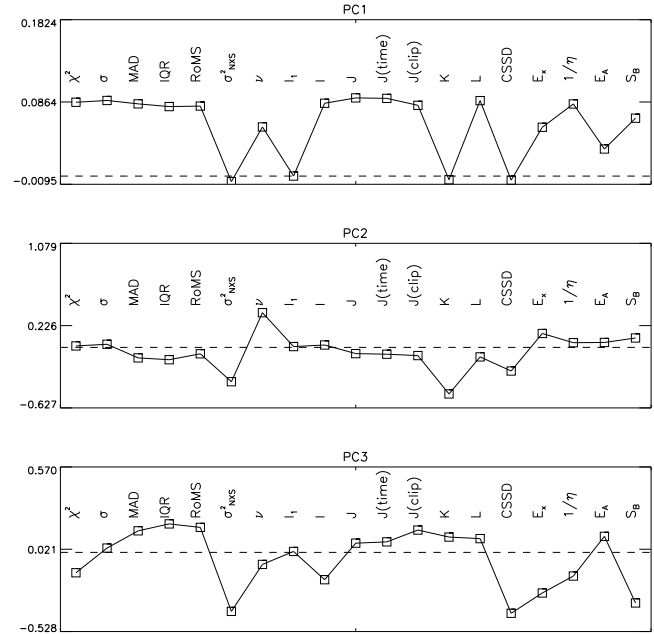}
 \caption{The first three principal components in the Kr data set
 (Section~\ref{sec:krasnoyarsk}). The dashed line indicates zero
 contribution of an index to the PC. Similar plots for the other data sets
 may be found at \url{http://scan.sai.msu.ru/kirx/var_idx_paper/}}
 \label{fig:PC}
\end{figure}

We consider the first three PCs.
For the Kr data set we find that PC1 is responsible for 56.5\% of the data variance, 
PC2 for 8.2\% and PC3 for 7.1\%.
The distribution of variance between the first principal
components for the other test data sets is very similar. 
The admixture coefficients corresponding to 
the first principal components  are presented in Fig.~\ref{fig:admixture}.
Variable objects tend to have large positive values of $\alpha_{1}$, while
they may have any $\alpha_{2}$ and $\alpha_{3}$ values. This suggests that
most of the information related to variability in general is encoded in PC1.
The components PC2 and PC3 may encode lightcurve characteristics that differ
for different variability types.
Fig.~\ref{fig:PC} presents the relative contribution of the variability
indices to the first three PCs. While many scatter and
correlation-based indices provide comparable contribution to PC1, the
indices $l_1$ and $\mathcal{E}_\mathcal{A}$ contribute less and the
contribution of $K$, CSSD, $\sigma_{\rm NXS}^2$ is near zero.
PC1 is dominated by the indices that generally perform better in identifying
variable objects (Section~\ref{sec:resdisc}).
The indices $l_1$, $S_B$, $\chi_{\rm red}^2$ contribute the most to PC2
while $K$, $v$ and $\sigma_{\rm NXS}^2$ dominate PC3.

\begin{figure}
 \centering
 \includegraphics[width=0.33\textwidth]{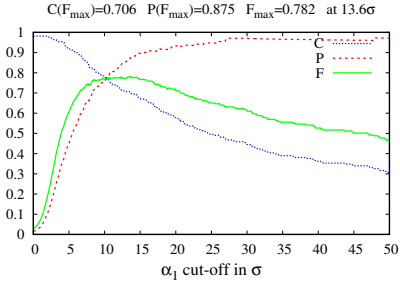}

 \caption{Variable star selection completeness (C),
purity (P) and $F_1$-score (F; see Section~\ref{sec:comparison_tech}) as a function of selection threshold for the
admixture coefficient $\alpha_{1}$ used as a composite variability index (Section~\ref{sec:pca})
computed for the Kr data set (Section~\ref{sec:krasnoyarsk}). Similar plots for the other data sets
 may be found online at \url{http://scan.sai.msu.ru/kirx/var_idx_paper/}}
 \label{fig:CPFPCAsample}
\end{figure}

The admixture coefficient $\alpha_{1}$ may be used as a
composite variability index since all variable objects tend to have large
positive values of $\alpha_{1}$ (Fig.~\ref{fig:admixture}).
It reaches the value of $F_1 = 0.659$ (Fig.~\ref{fig:CPFPCAsample}) 
and $R = 0.995$
on par with the best variability indices for this field
(Table~\ref{tab:FRreal}), but does not provide an improvement
over them. One possible use of $\alpha_{1}$ is to investigate a new
data set for which it is not known a~priori which variability indices 
are most suitable. In this case, one could compute multiple indices and perform
the PCA of them. The coefficient $\alpha_{1}$ is by construction one 
of the best variability indices (that captures most of variability-related information)
for this particular data set. 

\subsection{Limitations of the indices as variability indicators}

Besides the random errors (caused by the background and photon
noise\footnote{Scintillation noise may also contribute significantly to 
random errors in ground-based photometric observations if conducted with short exposures 
and small telescopes \citep[e.g.][]{2012A&A...546A..41K}.}) that are usually easy to estimate, 
photometric measurements are subject to
systematic error (due to atmospheric and instrumental variation) that are
hard to quantify. Since the overall measurement errors are not accurately
known, it is not possible to apply the $\chi^2$ test (Section~\ref{sec:chi2}) 
to select (non-)variable objects.
The absence of accurate error estimates can be substituted with the
assumptions that {(i)}\,the majority of field stars are non-variable and
{(ii)}\,stars of similar brightness in a given field are measured with 
about the same photometric accuracy. 
If these assumptions hold, the field stars may be used to measure
the actual accuracy of a given set of photometric observations. 
The variability indices (Section~\ref{sec:indices}) can be used to select
objects showing larger-than-expected brightness variations.

Since source extraction is not perfect, in practice there are some objects
measured with far worse accuracy than the majority, breaking the assumption
{(ii)} above. The source extraction problems may be caused by blending
and image artefacts.
Neither scatter nor correlation-based indices are effective in
distinguishing true variable objects from the ones with corrupted
photometry, which ultimately limits the usefulness of variability
indices.
The number of bad measurements in a photometric data set has a
higher impact on the efficiency of variability search than the choice of a
particular variability index. This is illustrated by comparison of
variability search results in the data sets TF1/TF2 (Section~\ref{sec:kpsobs}) and Kr
(Section~\ref{tab:testdatasummary}) obtained with similar equipment. The Kr
data set in which bad measurements are aggressively removed
provides systematically higher $F_{1~\rm max}$ scores 
than the TF1/TF2 data sets in which no flagging of bad
measurements is applied (Table~\ref{tab:FRreal}). The cost of removing
`suspicious' measurements that may be corrupted due to blending is that
one may lose some variable stars that are blended, but have sufficiently high variability
amplitude to be detected. The efficiency of variable star search with
variability indices is determined by the ability to identify and discard bad
measurements at the source extraction stage or assign appropriately high
error bars to such measurements (and then use a variability index that takes
errorbars into account, see Table~\ref{tab:indexsummary}).

By computing the indices one may pre-select candidate variables from
a photometric data set reducing the initial number of considered objects 
typically
by an order of magnitude.
An index-based selection of candidates should be followed by a more sophisticated 
analysis such as period search and visual inspection of lightcurves and images to
distinguish true variables from badly measured objects.

\subsection{How to select a cut-off value?}
\label{sec:cutoffvalue}

The cut-off value, $a$, for variable objects selection, which 
provides a balance between the selection completeness and
false positive rate (maximizing the $F_1$-score;
Section~\ref{sec:comparison_tech}) varies greatly between
indices and data sets (Fig.~\ref{fig:CPFsample}). To select $a$ for a new
variability survey one may use known variable stars covered by that survey.
One would often tolerate a large number of false-candidates in
favour of a more complete variable objects selection, so a threshold set by
maximizing the $F_1$-score (while being useful for comparing variability indices
with each other) 
may be considered too high in practice. Instead, it is possible to search for the
value of $a$ maximizing
\begin{equation}
F_\beta = (1+\beta^2) (C \times P) / (C + \beta^2 P),
\end{equation}
where the parameter $\beta>0$ determines how much importance we attach to
completeness, $C$, relative of purity, $P$. For the test data sets described
in Section~\ref{sec:testdata}, values of $\beta$ as high as 50 are needed to 
have most of the known variables selected (with the majority of indices) 
above the cut-off limit that maximizes $F_\beta$.

For any variability index, the distributions of index values
for variable and non-variable objects inherently intersect since 
{(i)}\,there is no lower limit on the possible amplitude of variability
and {(ii)}\,there are often some objects with corrupted
measurements resulting in elevated variability index values for them.
The value of $a$ should be chosen based on the false candidate rate that
can be practically handled at the post-processing stage. For example, only a small number of
false candidates are acceptable if selection based on variability indices is
immediately followed by a visual inspection. A larger number of false
candidates can be accepted if variability index-based selection is followed
by a period search. If no list of known variables is available for the new
survey data, one may start by setting, for example, $a=3$ and gradually
lowering the cut-off level until the number of false detections becomes
unacceptable.

\section{Conclusions}
\label{sec:conclusions}

We compare 18 variability indices quantifying the overall
scatter and/or degree of correlations between consecutive measurements in a
lightcurve. The ability of these indices to distinguish variable stars from 
non-variable ones is tested on seven data sets collected with various ground-based 
telescopes and on simulated data incorporating actual lightcurves of non-variable
objects as realistic models of photometric noise. We apply the PCA in search for an optimal combination of multiple
variability indices.

We find that correlation-based indices are 
more efficient in selecting variable objects than the scatter-based indices
for data sets containing hundreds of measurement epochs or more.
The indices $1/\eta$, $L$, ${\rm MAD}$ and ${\rm IQR}$ perform better than
others in selecting candidate variables from data sets affected by outliers.
We suggest using the $1/\eta$ index together with the ${\rm IQR}$ as the
pair of indices applicable to a wide variety of survey strategies
and variability types.
The indices $1/\eta$ and ${\rm IQR}$ provide stable high
performance, albeit not always the highest one for each of the investigated data sets.
However, the overall quality of a photometric data set including the percentage of outlier
measurements and number of badly measured objects
has a higher impact on the efficiency of variability search than the choice of a specific
(set of) variability index(es).

Another efficient approach to variability detection is to
compute many scatter- and correlation-based variability indices and 
perform the PCA over them. 
The admixture coefficient of the first principal component can be used as 
the composite index most suitable for the particular data set under investigation.
This `composite index' will perform on par with the best
individual variability indices in this data set, but it requires no a~priori
knowledge of which indices are the best for the data set under investigation.

We also find that in practice, all the discussed variability indices 
as well as their combinations 
are not sufficient 
on their own
to automatically select variable stars from a large
set of lightcurves. The reason is that both variable and non-variable
stars are diverse groups: variables may have various lightcurve shapes,
while non-variable stars include both the majority
of objects displaying just noise and 
objects with photometry corrupted by nearby 
objects, cosmetic defects of a CCD, 
etc. The investigated indices cannot distinguish the badly measured objects from real
variables because the corrupted measurements not only increase the lightcurve scatter
(compared to a non-variable object of similar brightness), 
but may also mimic correlated variability (due to night-to-night 
seeing variations, drift of the object's image across a cosmetic defect and
so on).
If all causes of measurement corruption in a particular data set can be identified
and all such cases flagged at the source extraction stage, the discussed
variability indices may efficiently distinguish variable objects standing out
among the majority of non-variable stars.

At the same time, the variability indices are perfectly suitable to solve
the inverse problem: identify well-measured constant stars in a photometric
data set. The list of well-measured non-variable stars may be useful 
as photometric standards for calibration or targets for a search of
variations not intrinsic to these objects such as microlensing
events, occultations of stars by distant Solar system objects, etc.

The data sets used to test the variability indices were searched for variable objects
previously. Despite that, we were able to identify 124 new variable stars
during the tests. This highlights the fact that variability search
techniques originally used to investigate the data sets can be improved by
the application of the multiple variability indices tested here.
The information about the new variables is summarized in Table~\ref{tab:newvar}
and their lightcurves are presented in Fig.~\ref{fig:newvarlightcurves}.
The variability types are assigned according to the GCVS
system\footnote{\url{http://www.sai.msu.su/gcvs/gcvs/iii/vartype.txt}}
\citep{2009yCat....102025S} and high-amplitude $\delta$~Scuti/SX~Phoenicis stars
are indicated as HADS.

\section*{Acknowledgments}

We thank Dr.~Laurent Eyer, Dr.~Ioannis Georgantopoulos, 
and Dr.~Ilya Pashchenko for critically reading the manuscript, 
Nikolas Laskaris for the discussion of algorithms performance,
Dr.~Valerio Nascimbeni and Dr.~Luigi Bedin for providing additional
data for our technical tests and the entire Hubble Source Catalog 
\citep[HSC;][]{2016AJ....151..134W}
team for illuminating discussions and HSC data.
KVS thanks Dr.~Boris Safonov and Dr.~Dmitry Chulkov for a discussion of scintillation-noise effects on photometry
and Dmitry Litvinov for advice on how to improve the abstract. 
We acknowledge financial support by the European Space Agency (ESA) under
the `Hubble Catalog of Variables' programme, contract no.\,4000112940. 
The work was supported by Act 211 Government of the Russian
Federation, contract no.~02.A03.21.0006 and by the Ministry of Education and
Science of the Russian Federation (the basic part of the state assignment,
registration number~01201465056). 
KVS, SVA and NNS are supported by the Russian Foundation for Basic Research grant 13-02-00664. 
This publication makes use of data products from the AAVSO
Photometric All Sky Survey (APASS). Funded by the Robert Martin Ayers
Sciences Fund and the National Science Foundation.
This research has made use of the International Variable Star Index (VSX)
database, operated at AAVSO, Cambridge, Massachusetts, USA.
This research has made use of the VizieR catalogue access tool, CDS,
Strasbourg, France. The original description of the VizieR service is
presented by \cite{2000A&AS..143...23O}.
This paper makes use of data from the DR1 of the WASP data \citep{2010A&A...520L..10B} as provided by the WASP consortium,
and the computing and storage facilities at the CERIT Scientific Cloud, reg. no. CZ.1.05/3.2.00/08.0144
which is operated by Masaryk University, Czech Republic.
This research has made use of NASA's Astrophysics Data System.

\footnotesize{
 \bibliographystyle{mn2e}
 \bibliography{vid}
}

\clearpage
\onecolumn

\setcounter{table}{3}
\begin{table*}
 \caption{New variable stars found in the test data}
 \label{tab:newvar}
 \begin{tabular}{c l c c c c c}
 \hline
 Name & Alias & $\upalpha_{\text{J2000}}$ $\updelta_{\text{J2000}}$ & Mag. range & Type & Period ($d$) & Epoch \\
 \hline
ogle\_17707  &  LMC\_SC20\_17707           &  05:44:59.79 $-$70:53:45.1  & 17.85-17.95\,I &  SR     &  86  &  max 2451164.798  \\
ogle\_33977  &  LMC\_SC20\_33977           &  05:45:07.10 $-$70:38:57.3  &  19.00-19.3\,I &  E:/CEP:  &  8.220  &  min 2450842.770  \\
ogle\_14141  &  LMC\_SC20\_14141           &  05:45:07.95 $-$70:56:55.9  & 17.65-17.75\,I &  SR     &  34.5  &  max 2450726.854  \\
ogle\_32793  &  LMC\_SC20\_32793           &  05:45:13.19 $-$70:39:12.2  &   17.8-18.1\,I &  GCAS   &   &   \\
ogle\_7022   &  LMC\_SC20\_7022            &  05:45:15.24 $-$71:04:59.4  & 17.35-17.50\,I &  SRD:   &  27.4  &  max 2451516.701  \\
ogle\_25034  &  LMC\_SC20\_25034           &  05:45:15.53 $-$70:46:23.8  &   17.3-17.4\,I &  SRD:   &  41.3  &  max 2451478.820  \\
ogle\_32597  &  LMC\_SC20\_32597           &  05:45:23.34 $-$70:39:05.4  &   16.9-17.2\,I &  L:     &   &   \\
ogle\_17541  &  LMC\_SC20\_17541           &  05:45:24.40 $-$70:53:53.0  & 16.80-16.95\,I &  L:/SR  &   &   \\
ogle\_43681  &  LMC\_SC20\_43681           &  05:45:29.65 $-$70:30:18.5  &   16.7-16.8\,I &  LB:    &   &   \\
ogle\_72706  &  LMC\_SC20\_72706           &  05:45:37.68 $-$70:53:39.3  & 16.55-16.65\,I &  L:/BE:  &   &   \\
ogle\_83546  &  LMC\_SC20\_83546           &  05:45:53.41 $-$70:41:45.0  & 16.85-16.95\,I &  SRD:   &  64.3  &  max 2450950.547  \\
ogle\_72940  &  LMC\_SC20\_72940           &  05:46:05.99 $-$70:55:20.0  &   18.8-19.1\,I &  SR     &  22.5  &  max 2451599.752  \\
ogle\_63585  &  LMC\_SC20\_63585           &  05:46:11.65 $-$71:03:24.1  &   17.4-17.6\,I &  L:/BE:  &   &   \\
ogle\_79972  &  LMC\_SC20\_79972           &  05:46:15.96 $-$70:45:19.2  & 17.70-17.85\,I &  SR     &  71.1  &  max 2451610.737  \\
ogle\_141796 &  LMC\_SC20\_141796          &  05:46:31.30 $-$70:36:54.9  & 16.85-17.00\,I &  BE:    &   &   \\
ogle\_178524 &  LMC\_SC20\_178524          &  05:47:02.75 $-$70:50:33.0  & 18.15-18.40\,I &  SR     &  78.7  &  max 2451628.671  \\
ogle\_178313 &  LMC\_SC20\_178313          &  05:47:12.07 $-$70:48:30.2  &   17.9-18.1\,I &  BY/EB  &  3.762  &  max 2451238.610  \\
ogle\_169150 &  LMC\_SC20\_169150          &  05:47:28.80 $-$70:59:52.8  &   17.8-18.0\,I &  SR     &   &   \\
ogle\_195557 &  LMC\_SC20\_195557          &  05:47:31.53 $-$70:31:52.5  & 16.65-16.75\,I &  LB:    &   &   \\
ogle\_195589 &  LMC\_SC20\_195589          &  05:47:33.05 $-$70:34:06.5  & 17.75-17.95\,I &  SR     &  26.0  &  max 2451189.779  \\
w1\_03348    &  2MASS J16464503-4548311  &  16:46:45.04 $-$45:48:31.1  & 17.45-17.65\,I &  L:     &    &   \\
w1\_02905    &  B1.0 0441-0525678        &  16:47:22.89 $-$45:49:25.6  & 12.64-12.72\,I &  L:     &    &   \\
66oph\_00151 &  B1.0 0955-0320170        &  17:56:05.59 $+$05:32:57.7  & 15.35-15.65\,V &  EW     &  0.439073  &  min 2443195.59  \\
66oph\_00509 &  B1.0 0953-0319502        &  17:56:40.00 $+$05:21:15.8  &   14.4-14.5\,V &  EB     &  0.716296  &  min 2443262.54  \\
66oph\_00554 &  B1.0 0953-0319763        &  17:56:54.61 $+$05:19:42.1  &   14.0-14.2\,V &  LB     &   &   \\
66oph\_21457 &  B1.0 0954-0321246        &  17:57:16.72 $+$05:26:15.8  &   16.4-17.0\,V &  EW     &  0.331902  &  min 2443272.41  \\
66oph\_01548 &  B1.0 0944-0313124        &  17:57:30.81 $+$04:27:55.8  & 13.65-14.0:\,V &  EA     &  2.93797  &  min 2442922.49  \\
66oph\_00416 &  B1.0 0953-0323577        &  18:00:09.95 $+$05:23:32.4  &  14.35-14.5\,V &  EB     &  0.880076  &  min 2444455.30  \\
tf1\_2696    &  B1.0 1411-0351209        &  20:25:47.04 $+$51:09:36.5  &   13.5-13.9\,I &  SR:    &   &   \\
tf1\_10844   &  B1.0 1395-0356332        &  20:30:04.59 $+$49:33:40.1  & 13.30-13.34\,I &  BY     &  13.428  &  max 2456141.320  \\
tf1\_11332   &  2MASS 20301843+5018158   &  20:30:18.43 $+$50:18:15.9  & 12.55-12.80\,I &  EA     &  4.098  &  min 2456167.225  \\
tf1\_12827   &  B1.0 1400-0361680        &  20:31:01.94 $+$50:03:07.2  & 11.67-11.74\,I &  L:     &    &   \\
tf1\_12884   &  2MASS 20310375+5106588   &  20:31:03.76 $+$51:06:58.8  & 15.20-15.40\,I &  EW     &  0.321647  &  min 2456131.286  \\
tf1\_14712   &  B1.0 1395-0357521        &  20:31:59.76 $+$49:31:18.3  & 15.70-16.30\,I &  L:     &   &   \\
tf1\_16769   &  B1.0 1407-0362160        &  20:33:01.76 $+$50:46:06.7  & 13.45-13.73\,I &  EB     &  0.441056  &  min 2456131.327  \\
kr\_77163    &  B1.0 1423-0521396        &  22:42:32.35 $+$52:21:21.8  & 15.55-15.76\,V &  BY:    &  0.7351  &  max 2456181.2  \\
and1\_20086  &  B1.0 1395-0471170        &  22:42:49.50 $+$49:31:08.5  &  11.2-11.32\,V &  EW     &  0.8792886  &  min 2455963.19  \\
kr\_37961    &  B1.0 1425-0519635        &  22:43:21.89 $+$52:30:51.9  & 15.25-15.37\,V &  EB     &  0.59755  &  min 2456179.172  \\
and1\_19739  &  B1.0 1369-0523951        &  22:43:31.66 $+$46:59:51.9  &   12.6-13.2\,V &  EA:    &  1.42962  &  min 2455948.144  \\
kr\_17020    &  B1.0 1414-0463241        &  22:43:38.70 $+$51:29:39.9  & 15.40-16.00\,V &  EA     &  0.91889  &  min 2456175.229  \\
kr\_66856    &  B1.0 1421-0514213        &  22:43:47.96 $+$52:08:47.6  & 14.15-14.35\,V &  EA     &  3.764  &  min 2456181.067  \\
kr\_10405    &  B1.0 1412-0461724        &  22:44:41.85 $+$51:13:59.5  & 13.88-13.92\,V &  DSCT   &  0.107088  &  max 2456181.067  \\
kr\_83071    &  B1.0 1418-0488970        &  22:44:51.41 $+$51:52:56.5  & 12.62-12.70\,V &  LB     &    &   \\
and1\_24872  &  B1.0 1432-0496857        &  22:45:01.06 $+$53:14:16.5  &  12.2-12.7:\,V &  LB     &   &   \\
kr\_43307    &  B1.0 1428-0540643        &  22:45:08.66 $+$52:52:22.4  & 13.05-13.08\,V &  DSCT   &  0.086931  &  max 2456174.089  \\
kr\_05854    &  B1.0 1410-0462570        &  22:45:19.32 $+$51:03:03.7  & 14.25-14.31\,V &  GDOR   &  0.41078  &  max 2456173.195  \\
and1\_24286  &  B1.0 1409-0465793        &  22:45:22.22 $+$50:58:17.8  &   12.1-12.6\,V &  LB     &   &   \\
kr\_28162    &  B1.0 1419-0487936        &  22:45:25.85 $+$51:58:00.5  & 13.87-13.91\,V &  DSCT   &  0.117360  &  max 2456174.266  \\
kr\_29944    &  B1.0 1420-0501500        &  22:45:54.19 $+$52:02:38.1  & 12.57-12.59\,V &  DSCT   &  0.054953  &  max 2456181.301  \\
kr\_91476    &  B1.0 1431-0520266        &  22:45:55.06 $+$53:06:22.1  & 13.18-13.35\,V &  LB     &    &   \\
kr\_46864    &  B1.0 1427-0539839        &  22:45:55.73 $+$52:43:57.9  & 14.52-14.57\,V &  DSCT   &  0.093029  &  max 2456179.202  \\
kr\_71058    &  B1.0 1424-0522460        &  22:45:56.73 $+$52:24:33.3  & 15.73-15.98\,V &  EW     &  0.32395  &  min 2456179.263  \\
kr\_20441    &  B1.0 1416-0479986        &  22:46:07.33 $+$51:39:44.4  & 12.25-12.40\,V &  LB     &    &   \\
kr\_66835    &  B1.0 1408-0471564        &  22:46:07.37 $+$50:48:58.5  & 14.09-14.17\,V &  EW     &  0.35198  &  min 2456171.143  \\
kr\_77060    &  B1.0 1417-0490596        &  22:46:09.32 $+$51:42:26.4  & 15.20-15.85\,V &  EA     &  1.4150  &  min 2456179.303  \\
kr\_01539    &  B1.0 1408-0471596        &  22:46:09.35 $+$50:52:33.6  & 14.33-14.39\,V &  DSCT   &  0.181908  &  max 2456181.385  \\
and1\_23835  &  B1.0 1374-0578230        &  22:46:09.39 $+$47:27:39.3  &  12.24-12.8\,V &  LB     &   &   \\
kr\_12545    &  B1.0 1413-0473694        &  22:46:20.01 $+$51:20:19.8  & 14.62-14.80\,V &  LB     &    &   \\
kr\_15917    &  B1.0 1414-0466085        &  22:46:44.59 $+$51:28:47.1  & 14.02-14.06\,V &  DSCT   &  0.060571  &  max 2456173.233  \\
kr\_31044    &  2MASS 22465045+5205269   &  22:46:50.46 $+$52:05:27.0  & 15.24-15.37\,V &  EW     &  0.35722  &  min 2456181.15  \\
kr\_47989    &  B1.0 1426-0532761        &  22:47:13.65 $+$52:41:31.9  & 14.46-14.55\,V &  GDOR:  &  1.50:  &  max 2456181.05  \\
kr\_14653    &  B1.0 1414-0466538        &  22:47:16.41 $+$51:25:52.9  & 14.20-14.28\,V &  GDOR:  &  2.533  &  max 2456180.20  \\
kr\_55748    &  B1.0 1423-0526261        &  22:47:28.39 $+$52:22:44.3  & 12.10-12.22\,V &  LB     &    &   \\
kr\_48798    &  B1.0 1426-0533265        &  22:47:43.86 $+$52:39:59.7  & 15.04-15.15\,V &  BY:    &  1.646:  &  max 2456181.28  \\
 \end{tabular}
\end{table*}
\addtocounter{table}{-1}
\begin{table*}
 \caption{ continued}
 \label{tab:newvar}
 \begin{tabular}{c l c c c c c}
 \hline
 Name & Alias & $\upalpha_{\text{J2000}}$ $\updelta_{\text{J2000}}$ & Mag. range & Type & Period ($d$) & Epoch \\
 \hline
kr\_52054    &  B1.0 1425-0524424        &  22:48:26.87 $+$52:31:52.2  & 14.54-14.80:\,V &  EA     &    &   \\
kr\_22423    &  B1.0 1417-0492755        &  22:48:27.36 $+$51:45:46.4  & 13.48-13.51\,V &  DSCT   &  0.068565  &  max 2456171.137  \\
kr\_49316    &  B1.0 1426-0534324        &  22:48:48.79 $+$52:39:15.3  & 14.42-14.55\,V &  BY     &  2.234  &  max 2456205.13  \\
kr\_22913    &  B1.0 1417-0493245        &  22:49:02.03 $+$51:47:07.1  & 13.50-13.53\,V &  EW     &  0.255044  &  min 2456174.146  \\
kr\_25048    &  B1.0 1418-0493332        &  22:49:23.75 $+$51:52:15.4  & 13.74-13.78\,V &  DSCT   &  0.121500  &  max 2456175.095  \\
and1\_28521  &  B1.0 1426-0535417        &  22:50:00.00 $+$52:36:54.8  &   11.9-12.4\,V &  LB     &   &   \\
kr\_71053    &  B1.0 1420-0505962        &  22:50:47.92 $+$52:04:51.3  & 15.75-15.93\,V &  EW     &  0.48206  &  min 2456174.11  \\
kr\_68645    &  B1.0 1428-0546648        &  22:51:18.85 $+$52:49:18.8  & 15.43-15.60\,V &  EW     &  0.53633  &  min 2456175.11  \\
kr\_14338    &  B1.0 1414-0470583        &  22:51:49.59 $+$51:27:09.0  & 15.40-15.90\,V &  EA     &  0.95253  &  min 2456171.152  \\
kr\_11898    &  B1.0 1413-0479241        &  22:52:31.70 $+$51:21:31.3  & 13.43-13.46\,V &  DSCT   &  0.084288  &  max 2456159.311  \\
kr\_43786    &  B1.0 1428-0547936        &  22:52:31.83 $+$52:53:50.0  & 14.92-15.20\,V &  EA     &  1.4044  &  min 2456180.15  \\
kr\_05250    &  B1.0 1410-0468964        &  22:52:44.29 $+$51:05:12.1  & 12.17-12.19\,V &  DSCT   &  0.114447  &  max 2456171.139  \\
kr\_06442    &  B1.0 1411-0468958        &  22:52:46.95 $+$51:08:17.5  & 12.39-12.51\,V &  LB     &    &   \\
kr\_17485    &  B1.0 1415-0474346        &  22:53:10.52 $+$51:35:26.0  & 13.37-13.39\,V &  DSCT   &  0.071639  &  max 2456174.227  \\
kr\_06681    &  B1.0 1411-0469388        &  22:53:20.42 $+$51:08:47.5  & 13.36-13.40\,V &  DSCT   &  0.074128  &  max 2456173.159  \\
kr\_80746    &  B1.0 1430-0529127        &  22:53:21.22 $+$53:03:29.5  & 15.90-16.30\,V &  EA     &  0.48435  &  min 2456181.269  \\
kr\_68763    &  B1.0 1413-0479941        &  22:53:30.47 $+$51:20:02.6  & 15.41-15.52\,V &  DSCT   &  0.055500  &  max 2456179.308  \\
kr\_36496    &  B1.0 1429-0540150        &  22:53:35.71 $+$52:59:38.5  & 15.19-15.32\,V &  BY:    &  1.733  &  max 2456182.07  \\
kr\_74877    &  B1.0 1429-0540181        &  22:53:38.20 $+$52:58:46.3  & 16.05-16.22\,V &  EB     &  0.52761  &  min 2456180.147  \\
kr\_04524    &  B1.0 1410-0469644        &  22:53:41.52 $+$51:03:38.2  & 13.09-13.13\,V &  DSCT   &  0.068231  &  max 2456173.325  \\
kr\_54963    &  B1.0 1424-0530011        &  22:54:04.57 $+$52:26:15.1  & 15.02-15.14\,V &  DSCT   &  0.094157  &  max 2456173.222  \\
kr\_20503    &  B1.0 1417-0497638        &  22:54:11.29 $+$51:42:59.6  & 13.18-13.21\,V &  DSCT   &  0.062144  &  max 2456171.202  \\
kr\_20269    &  B1.0 1417-0497693        &  22:54:15.71 $+$51:42:22.4  & 14.38-14.50\,V &  EA     &  0.66044  &  min 2456180.285  \\
and1\_33002  &  B1.0 1398-0476733        &  22:54:30.93 $+$49:50:04.0  &  12.5-13.65\,V &  SR     &  90:  &  max 2456017.245  \\
kr\_67216    &  B1.0 1421-0523860        &  22:54:31.35 $+$52:06:14.8  & 14.83-14.91\,V &  DSCT   &  0.062844  &  max 2456177.256  \\
and1\_33250  &  B1.0 1436-0432148        &  22:54:33.11 $+$53:40:22.4  &  12.65-13.1\,V &  E:     &  1.5136  &  min 2455955.208  \\
and1\_33409  &  B1.0 1434-0440482        &  22:54:41.41 $+$53:29:11.9  &  12.63-13.0\,V &  EA     &  1.30331  &  min 2455869.245  \\
kr\_27255    &  B1.0 1419-0496162        &  22:54:43.15 $+$51:59:46.2  & 13.32-13.36\,V &  GDOR   &  2.85  &  max 2456183.07  \\
kr\_34551    &  B1.0 1430-0530571        &  22:54:52.44 $+$53:04:36.3  & 12.57-12.76\,V &  LB     &    &   \\
kr\_37294    &  B1.0 1429-0541415        &  22:54:55.72 $+$52:57:51.9  & 12.29-12.31\,V &  DSCT   &  0.035091  &  max 2456174.230  \\
kr\_28139    &  B1.0 1420-0509578        &  22:55:11.94 $+$52:01:49.2  & 14.36-14.40\,V &  EW     &  0.44403  &  min 2456179.300  \\
kr\_08347    &  B1.0 1412-0470958        &  22:55:44.63 $+$51:13:56.7  & 12.22-12.25\,V &  DSCT   &  0.107793  &  max 2456173.259  \\
kr\_38248    &  B1.0 1425-0531121        &  22:55:46.30 $+$52:34:29.4  & 12.39-12.42\,V &  DSCT   &  0.057451  &  max 2456159.231  \\
kr\_92507    &  B1.0 1431-0531790        &  22:56:44.75 $+$53:10:52.2  & 14.62-14.79\,V &  EW     &  0.50843  &  min 2456171.26  \\
kr\_41182    &  B1.0 1422-0536615        &  22:57:16.20 $+$52:15:53.3  & 14.53-14.58\,V &  GDOR   &  0.48395  &  max 2456153.282  \\
kr\_49209    &  B1.0 1426-0541925        &  22:57:18.92 $+$52:41:14.5  & 14.11-14.19:\,V &  EB     &  1.100:  &  min 2456179.39  \\
and1\_37044  &  B1.0 1405-0476718        &  22:58:26.06 $+$50:33:19.3  &  12.2-12.98\,V &  LB     &   &   \\
and1\_37967  &  B1.0 1416-0491257        &  22:59:21.59 $+$51:39:38.8  & 12.06-12.35\,V &  LB     &   &   \\
and1\_37854  &  B1.0 1381-0569982        &  22:59:24.58 $+$48:08:10.4  &  13.15-13.6\,V &  BY:    &  26  &  max 2456018.923  \\
and1\_37969  &  B1.0 1414-0476215        &  22:59:31.35 $+$51:29:09.9  & 10.48-10.76\,V &  LB     &   &   \\
and1\_38659  &  B1.0 1417-0501964        &  22:59:58.74 $+$51:43:57.8  &  12.41-12.7\,V &  EW:    &  0.457418  &  min 2455961.293  \\
and1\_18902  &  B1.0 1413-0485310        &  23:01:04.56 $+$51:20:48.6  & 10.91-11.12\,V &  SR     &  18.8  &  max 2456021.124  \\
and1\_18328  &  B1.0 1404-0488506        &  23:01:45.41 $+$50:26:57.9  & 12.52-13.0:\,V &  EA     &  3.5973  &  min 2455993.367  \\
and1\_17680  &  B1.0 1405-0479397        &  23:02:19.30 $+$50:31:01.2  &  13.57-13.9\,V &  EB     &  0.8137610  &  min 2456020.531  \\
and1\_15410  &  B1.0 1365-0495019        &  23:04:03.28 $+$46:32:10.5  & 11.55-11.85\,V &  SR     & 25.8  &  max 2456426.931  \\
and1\_15631  &  B1.0 1422-0541972        &  23:04:04.54 $+$52:17:03.4  &  11.7-12.15\,V &  LB     &   &   \\
and1\_15618  &  B1.0 1434-0448399        &  23:04:16.40 $+$53:29:44.8  & 11.76-11.96\,V &  HADS:  &  0.1335509  &  max 2456383.546  \\
and1\_14823  &  B1.0 1428-0558741        &  23:05:01.25 $+$52:49:36.8  &   13.2-13.7\,V &  EB     &  0.71946  &  min 2455951.183  \\
and1\_14065  &  B1.0 1396-0483812        &  23:05:09.40 $+$49:40:48.8  & 10.75-11.15\,V &  SR:    &  62  &  max 2456020.531  \\
and1\_12184  &  B1.0 1389-0486242        &  23:06:12.02 $+$48:57:18.5  &  9.92-10.32\,V &  LB     &   &   \\
and1\_13027  &  B1.0 1433-0484201        &  23:06:44.52 $+$53:23:57.1  &  12.2-12.85\,V &  SR:    &  70:  &  \\
and1\_11105  &  B1.0 1378-0608568        &  23:07:44.39 $+$47:51:44.1  & 11.44-11.75\,V &  LB     &   &   \\
and1\_07939  &  B1.0 1400-0490607        &  23:10:57.16 $+$50:03:03.4  &   10.7-11.5\,V &  SR:    &  60:  &   \\
and1\_07637  &  B1.0 1395-0490034        &  23:11:18.51 $+$49:33:20.7  & 11.75-12.14\,V &  EA     &  1.9246  &  min 2455969.161  \\
and1\_07465  &  B1.0 1429-0555802        &  23:12:00.10 $+$52:58:23.5  &   11.8-12.2\,V &  LB     &   &   \\
and1\_05237  &  B1.0 1409-0486197        &  23:13:48.68 $+$50:56:29.6  &  11.22-12.4\,V &  SR:    &  110  based on NSVS data  &   \\
and1\_05733  &  B1.0 1431-0545975        &  23:13:58.44 $+$53:10:01.1  & 11.27-11.46\,V &  SR     &  52.5  &  max 2455967.186  \\
and1\_03934  &  B1.0 1401-0500860        &  23:14:08.75 $+$50:11:58.5  & 10.53-10.83\,V &  SR:    &  20  &   \\
and1\_05292  &  B1.0 1419-0509999        &  23:14:09.71 $+$51:57:26.2  & 12.15-12.45\,V &  SR     &  40  &  max 2456003.171  \\
and1\_03638  &  B1.0 1396-0489427        &  23:14:52.69 $+$49:37:40.9  &  10.65-11.0\,V &  LB     &   &   \\
 \hline
 \end{tabular}
\end{table*}
\setcounter{table}{5}

\setcounter{figure}{1}
\begin{figure*}
 \centering
 \includegraphics[width=0.9\textwidth]{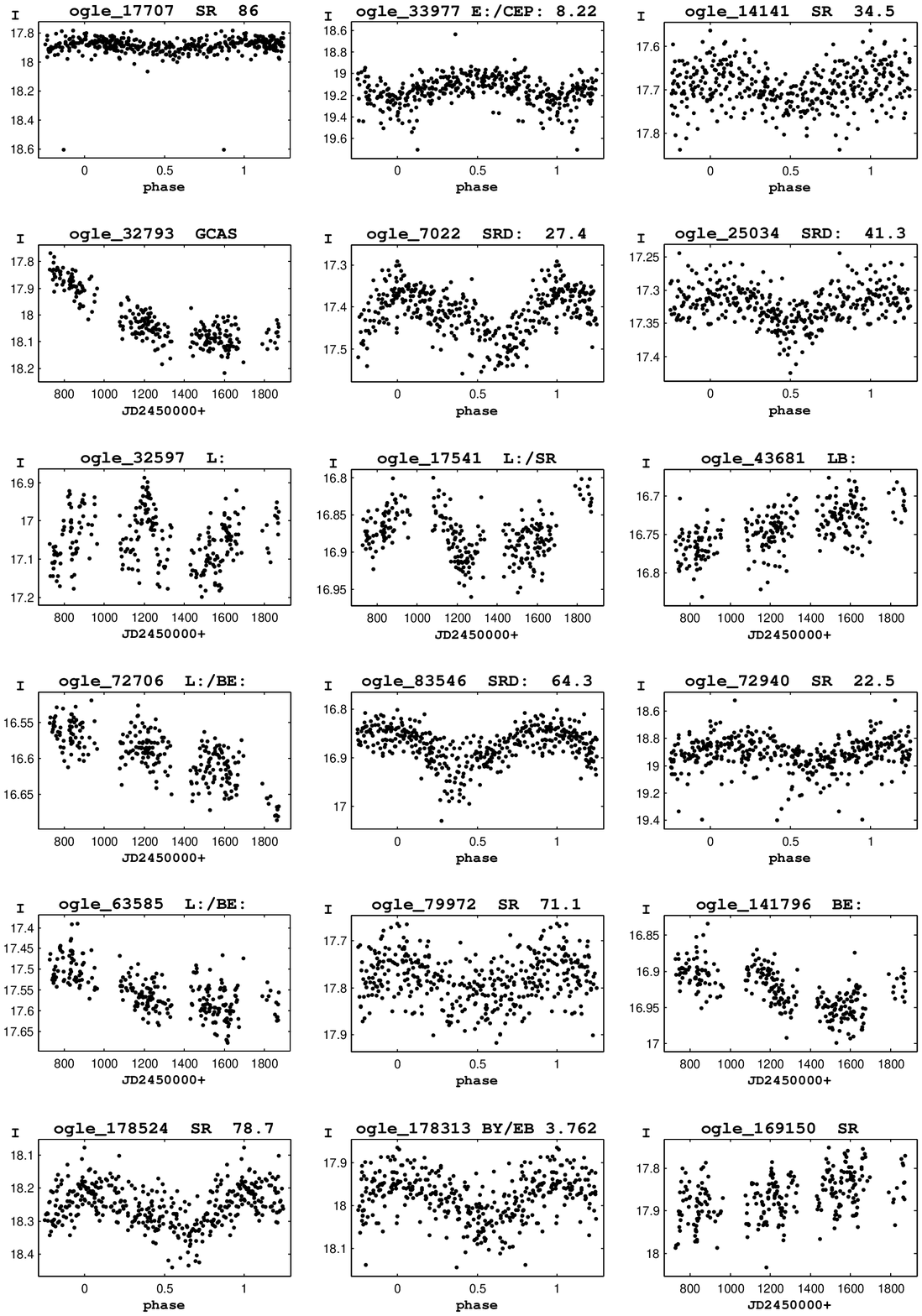}
 \caption{Lightcurves of new variable stars found in the test data sets. 
The magnitudes measured in a band indicated in the top-left corner of each
panel are plotted as a function of time (Julian day) for irregular or phase
for periodic variables. The title of each panel indicate the object
identifier in Table~\ref{tab:newvar}, its variability type and period in days (if applicable).}
 \label{fig:newvarlightcurves}
\end{figure*}
\addtocounter{figure}{-1}
\begin{figure*}
 \centering
 \includegraphics[width=0.9\textwidth]{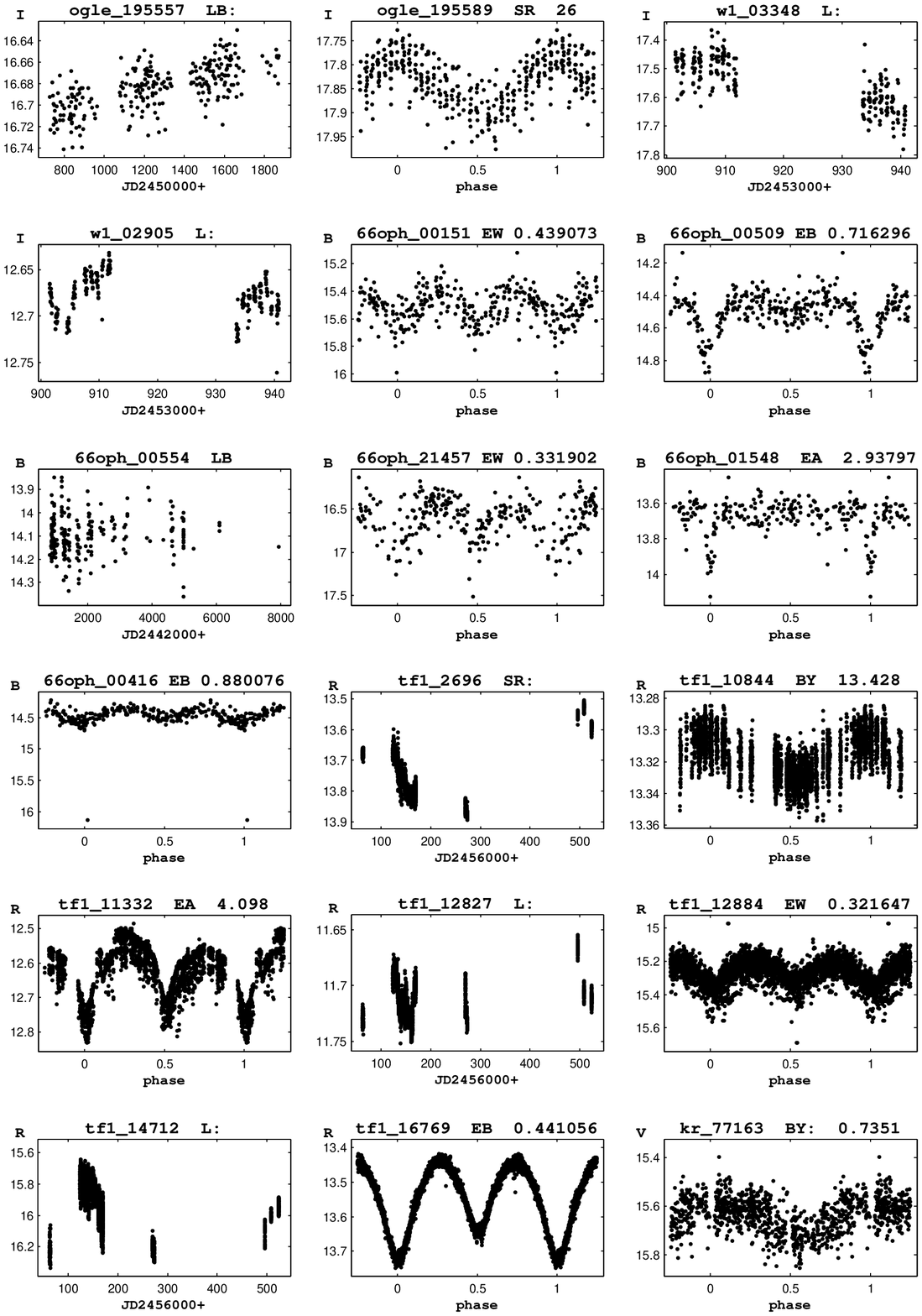}
 \caption{ continued. }
\end{figure*}
\addtocounter{figure}{-1}
\begin{figure*}
 \centering
 \includegraphics[width=0.9\textwidth]{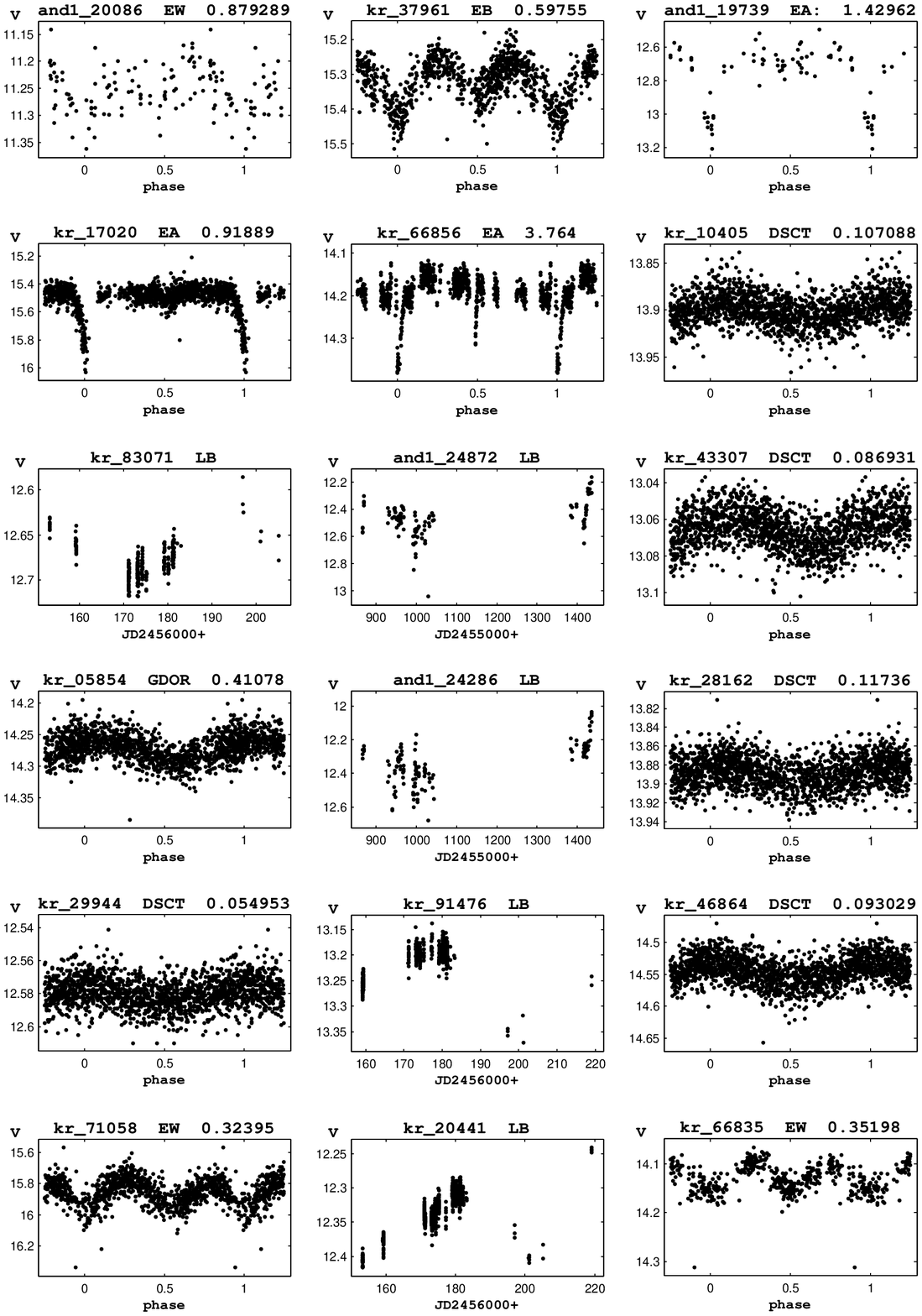}
 \caption{ continued. }
\end{figure*}
\addtocounter{figure}{-1}
\begin{figure*}
 \centering
 \includegraphics[width=0.9\textwidth]{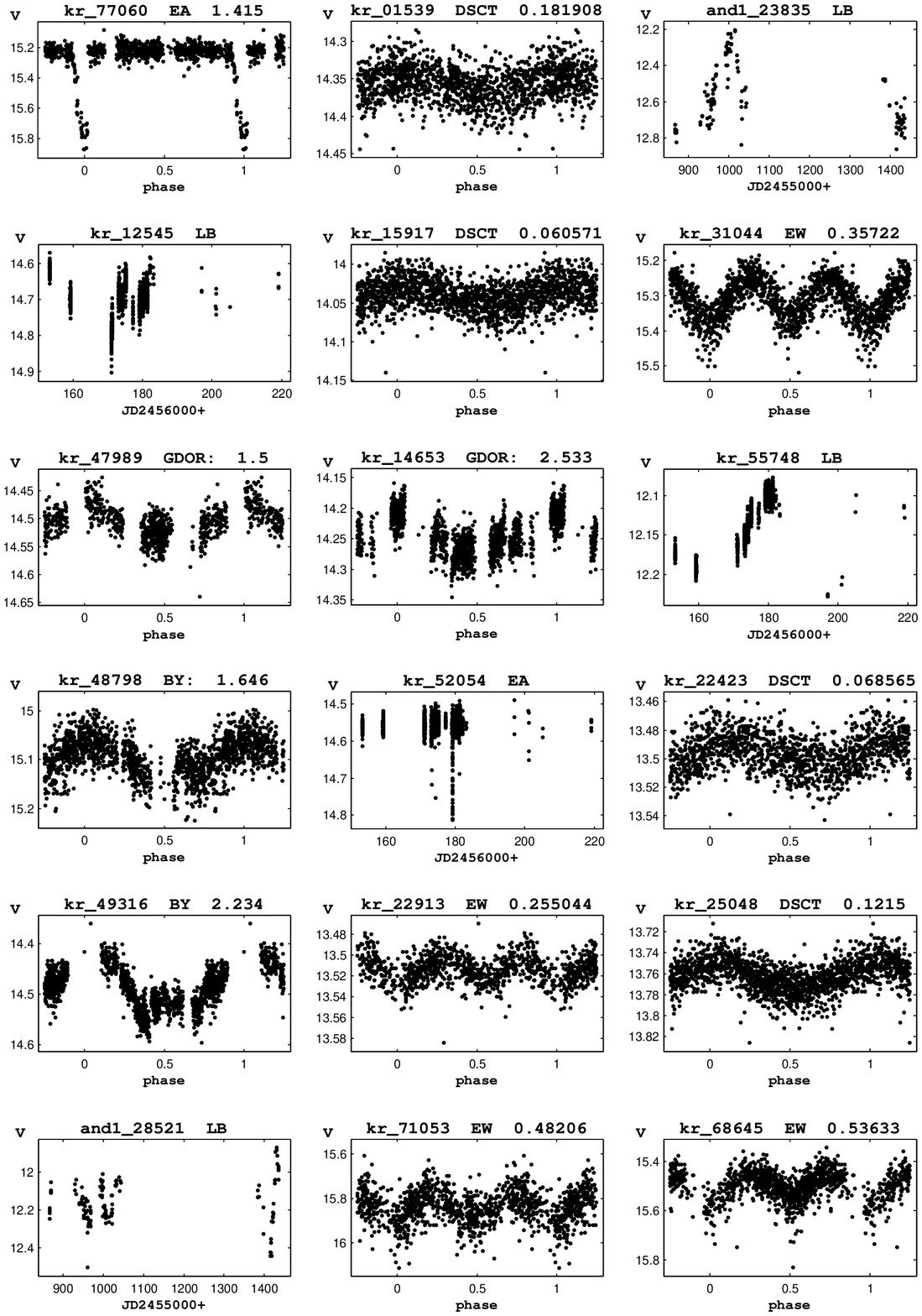}
 \caption{ continued. }
\end{figure*}
\addtocounter{figure}{-1}
\begin{figure*}
 \centering
 \includegraphics[width=0.9\textwidth]{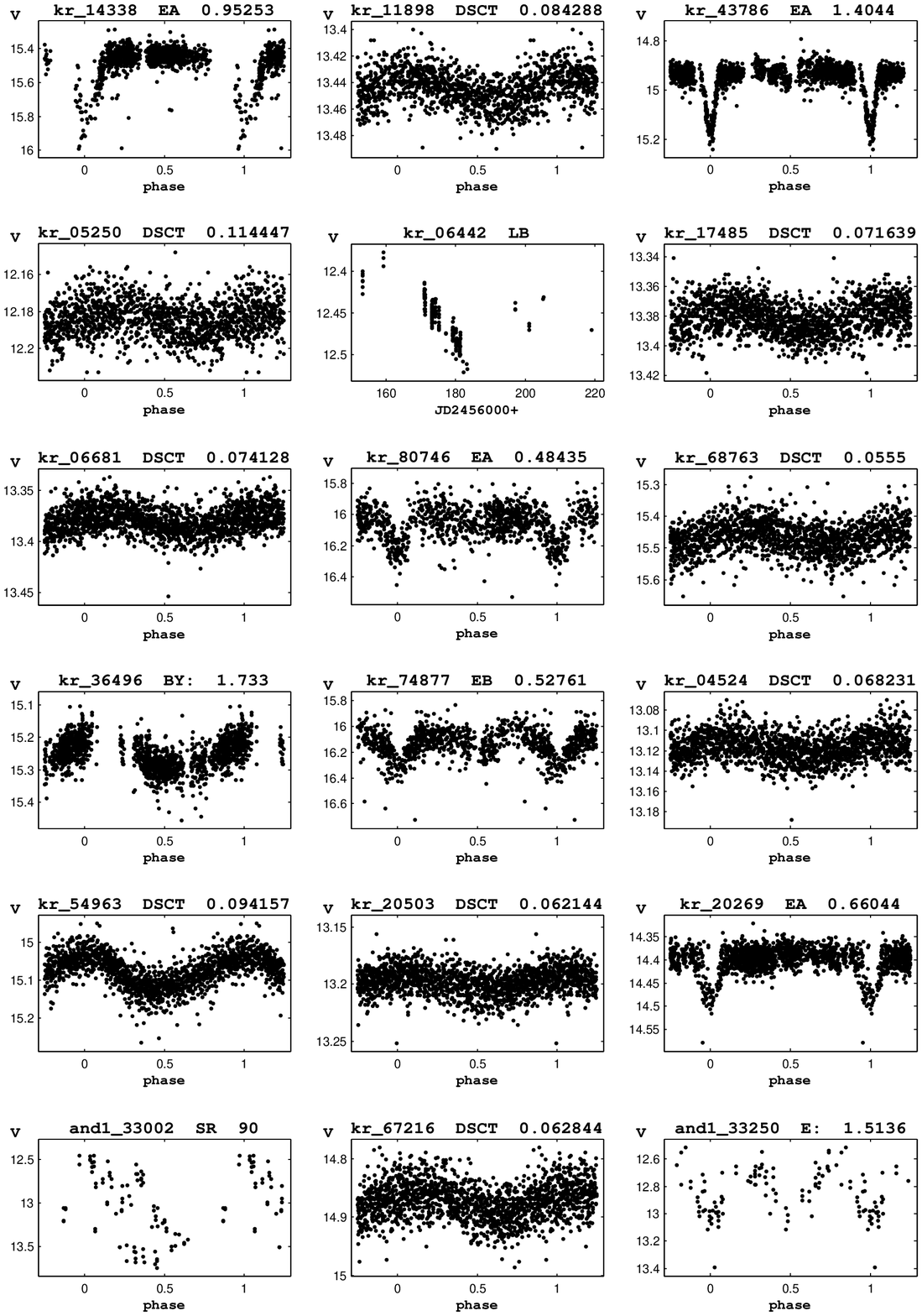}
 \caption{ continued. }
\end{figure*}
\addtocounter{figure}{-1}
\begin{figure*}
 \centering
 \includegraphics[width=0.9\textwidth]{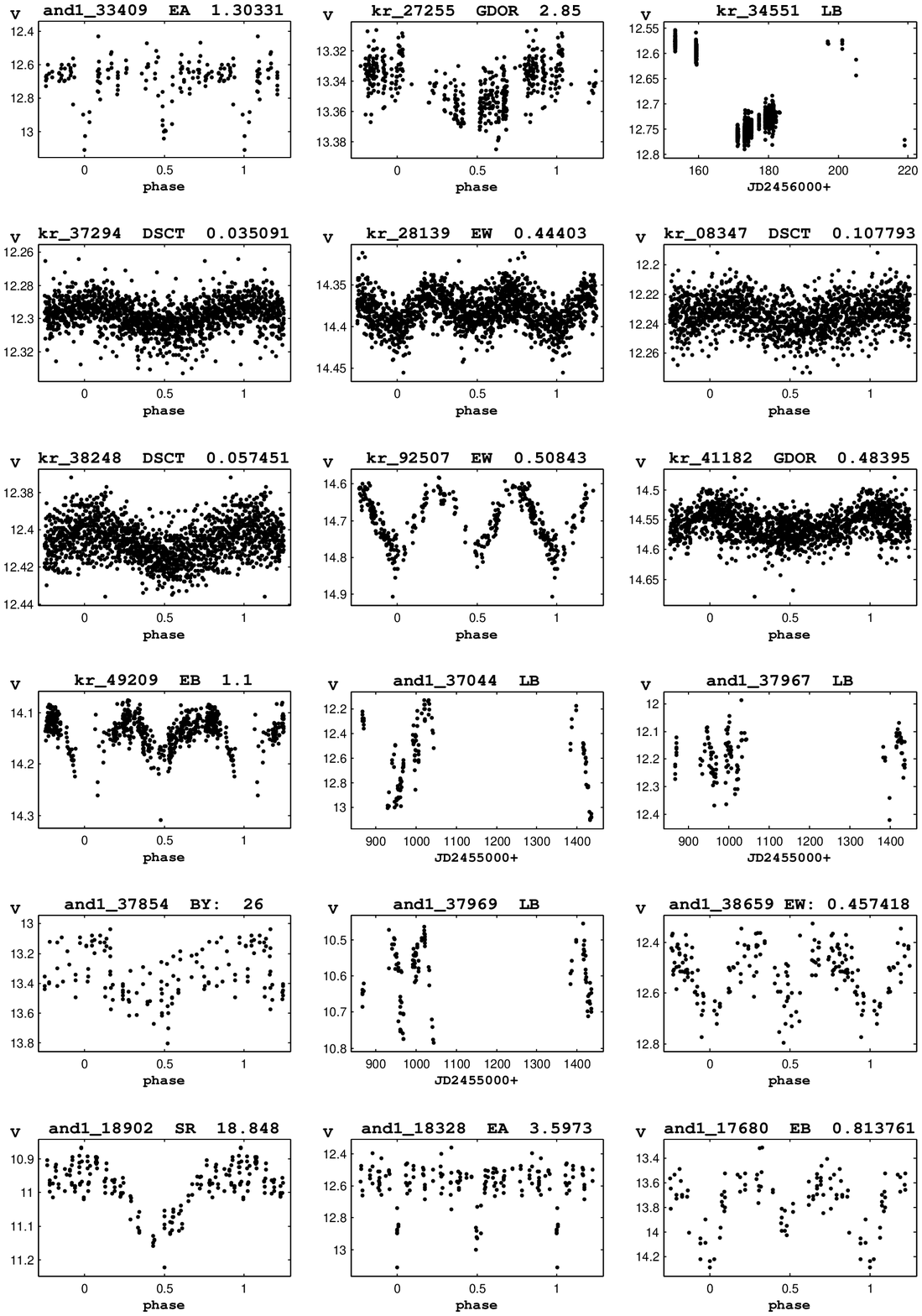}
 \caption{ continued. }
\end{figure*}
\addtocounter{figure}{-1}
\begin{figure*}
 \centering
 \includegraphics[width=0.9\textwidth]{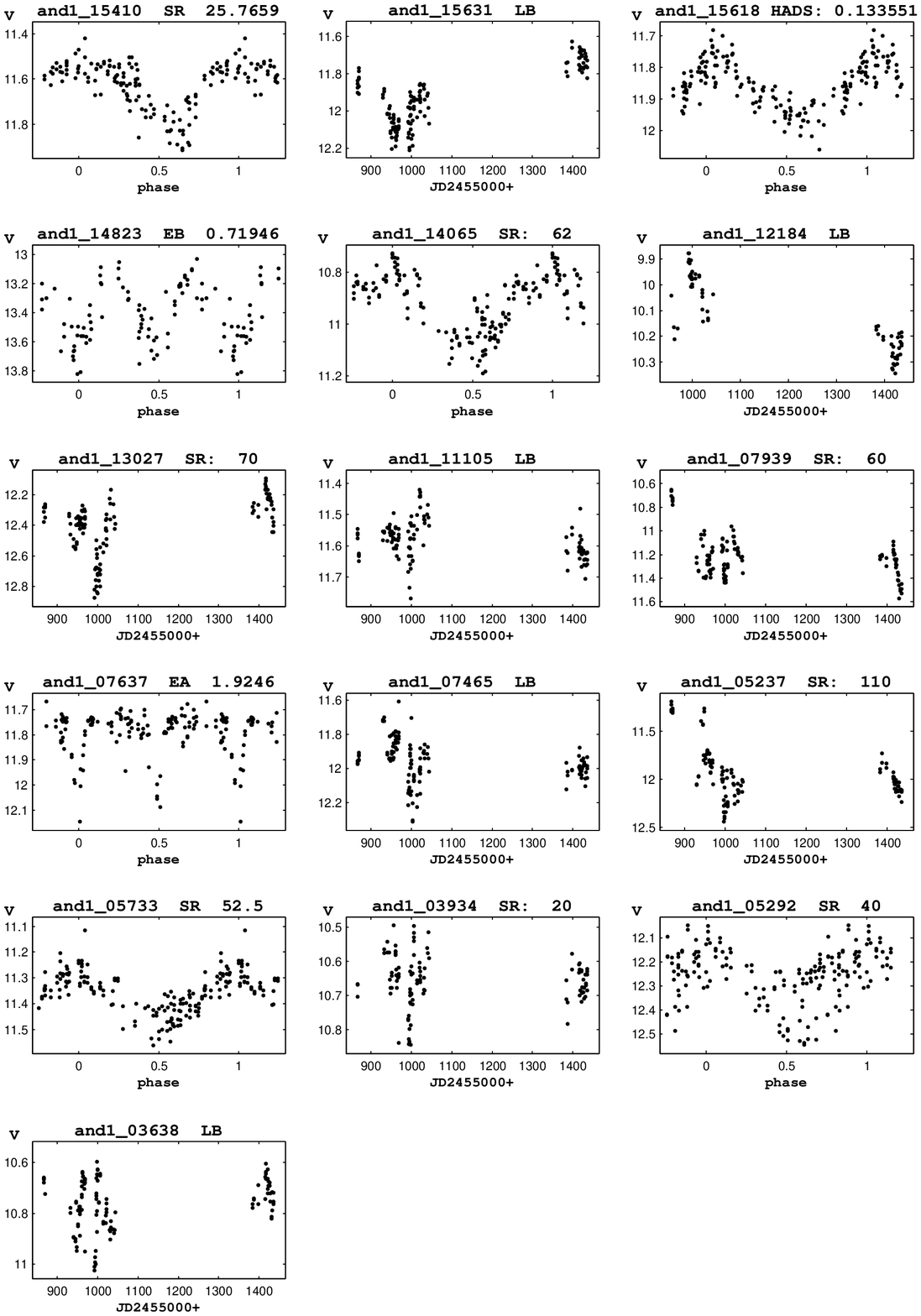}
 \caption{ continued. }
\end{figure*}
\setcounter{figure}{8}

\begin{table*}
   \caption{Performance of variability indices on the data sets with simulated periodic variability}
   \label{tab:FRsimperiodic}
   \begin{tabular}{c c@{~~}c c@{~~}c c@{~~}c c@{~~}c c@{~~}c c@{~~}c c@{~~}c cl}
   \hline\hline
                       & \multicolumn{2}{c}{TF1}  & \multicolumn{2}{c}{TF2}   & \multicolumn{2}{c}{Kr}       & \multicolumn{2}{c}{Westerlund\,1} & \multicolumn{2}{c}{And\,1} & \multicolumn{2}{c}{LMC\_SC20} & \multicolumn{2}{c}{66\,Oph} &        & \\
Index                  & $F_{1~\rm max}$ & $R$ & $F_{1~\rm max}$ & $R$ & $F_{1~\rm max}$ & $R$ & $F_{1~\rm max}$ & $R$ & $F_{1~\rm max}$ & $R$    & $F_{1~\rm max}$ & $R$  & $F_{1~\rm max}$ & $R$     & Sec. & Ref. \\
   \hline
\multicolumn{15}{c}{Scatter-based indices} \\
 $\chi_{\rm red}^2$     & 0.157 & 0.928   & 0.185 & 0.936   & 0.871 & 0.991   & 0.566 & 0.990   & 0.571 & 0.991   & 0.684 & 0.993   & 0.253 & 0.981  & \ref{sec:chi2}         & (a) \\
 $\sigma_w$             & 0.166 & 0.940   & 0.187 & 0.941   & 0.858 & 0.992   & 0.500 & 0.990   & 0.589 & 0.991   & 0.669 & 0.993   & 0.262 & 0.981  & \ref{sec:sigma}        & (b) \\
 ${\rm MAD}$            & 0.296 & 0.973   & 0.309 & 0.973   & 0.883 & 0.991   & 0.625 & 0.993   & 0.680 & 0.993   & 0.809 & 0.992   & 0.487 & 0.990  & \ref{sec:MAD}          & (c) \\
 ${\rm IQR}$            & 0.269 & 0.968   & 0.279 & 0.969   & 0.880 & 0.992   & 0.611 & 0.993   & 0.675 & 0.992   & 0.801 & 0.992   & 0.464 & 0.991  & \ref{sec:IQR}          & (d) \\
 ${\rm RoMS}$           & 0.215 & 0.957   & 0.237 & 0.954   & 0.882 & 0.991   & 0.654 & 0.993   & 0.626 & 0.992   & 0.768 & 0.991   & 0.437 & 0.993  & \ref{sec:RoMS}         & (e) \\
 $\sigma_{\rm NXS}^2$   & 0.023 & 0.243   & 0.022 & 0.196   & 0.065 & 0.737   & 0.042 & 0.599   & 0.061 & 0.758   & 0.069 & 0.775   & 0.346 & 0.990  & \ref{sec:sigmaxs}      & (f) \\
 $v$                    & 0.074 & 0.860   & 0.097 & 0.901   & 0.488 & 0.993   & 0.361 & 0.984   & 0.415 & 0.990   & 0.141 & 0.934   & 0.070 & 0.911  & \ref{sec:peaktopeak}   & (g) \\
\multicolumn{15}{c}{Correlation-based indices} \\
 $l_1$                  & 0.668 & 0.993   & 0.530 & 0.993   & 0.157 & 0.993   & 0.250 & 0.998   & 0.332 & 0.986   & 0.032 & 0.994   & 0.025 & 0.980  & \ref{sec:lag1autocorr} & (h) \\
 $I$                    & 0.164 & 0.926   & 0.193 & 0.935   & 0.877 & 0.991   & 0.286 & 0.993   & 0.641 & 0.991   & 0.354 & 0.994   & 0.097 & 0.989  & \ref{sec:welch}        & (i) \\
 $J$                    & 0.231 & 0.956   & 0.249 & 0.958   & 0.890 & 0.991   & 0.320 & 0.996   & 0.672 & 0.991   & 0.777 & 0.991   & 0.329 & 0.988  & \ref{sec:stetson}      & (j) \\
 $J({\rm time})$        & 0.240 & 0.959   & 0.249 & 0.958   & 0.891 & 0.991   & 0.345 & 0.996   & 0.685 & 0.992   & 0.279 & 0.996   & 0.081 & 0.992  & \ref{sec:stetsonTime}  & (k) \\
 $J({\rm clip})$        & 0.217 & 0.951   & 0.239 & 0.954   & 0.898 & 0.991   & 0.622 & 0.994   & 0.642 & 0.991   & 0.768 & 0.991   & 0.433 & 0.993  & \ref{sec:stetsonClip}  & (d) \\
 $L$                    & 0.272 & 0.964   & 0.274 & 0.961   & 0.877 & 0.991   & 0.323 & 0.996   & 0.712 & 0.992   & 0.796 & 0.991   & 0.495 & 0.994  & \ref{sec:stetson}      & (j) \\
 ${\rm CSSD}$           & 0.295 & 0.959   & 0.408 & 0.972   & 0.020 & 0.011   & 0.020 & 0.016   & 0.020 & 0.011   & 0.020 & 0.012   & 0.019 & 0.002  & \ref{sec:cssd}         & (l) \\
 $E_x$                  & 0.077 & 0.863   & 0.083 & 0.895   & 0.303 & 0.993   & 0.132 & 0.981   & 0.439 & 0.990   & 0.714 & 0.992   & 0.255 & 0.989  & \ref{sec:excursions}   & (m) \\
 $1/\eta$               & 0.674 & 0.992   & 0.625 & 0.988   & 0.859 & 0.991   & 0.242 & 0.997   & 0.315 & 0.985   & 0.031 & 0.994   & 0.023 & 0.974  & \ref{sec:vonneumann}   & (n) \\
 $\mathcal{E}_\mathcal{A}$& 0.012 & 0.970   & 0.040 & 0.988   & 0.168 & 0.978   & 0.021 & 0.946   & 0.025 & 0.689   & 0.019 & 0.539   & 0.019 & 0.897  & \ref{sec:EA}           & (o) \\
 $S_B$                  & 0.138 & 0.893   & 0.141 & 0.888   & 0.826 & 0.990   & 0.292 & 0.984   & 0.469 & 0.989   & 0.492 & 0.991   & 0.136 & 0.986  & \ref{sec:sb}           & (p) \\
   \hline
   \end{tabular}
   \renewcommand{\arraystretch}{1.0}
\vspace{-8pt}
\begin{flushleft}See the footnote in Table~\ref{tab:FRreal}.\end{flushleft}
\end{table*}

\begin{table*}
   \caption{Performance of variability indices on the data sets with simulated non-periodic variability}
   \label{tab:FRsimnonperiodic}
   \begin{tabular}{c c@{~~}c c@{~~}c c@{~~}c c@{~~}c c@{~~}c c@{~~}c c@{~~}c cl}
   \hline\hline
                       & \multicolumn{2}{c}{TF1}  & \multicolumn{2}{c}{TF2}   & \multicolumn{2}{c}{Kr}    & \multicolumn{2}{c}{Westerlund\,1} & \multicolumn{2}{c}{And\,1} & \multicolumn{2}{c}{LMC\_SC20} & \multicolumn{2}{c}{66\,Oph} &        & \\
Index                  & $F_{1~\rm max}$ & $R$ & $F_{1~\rm max}$ & $R$ & $F_{1~\rm max}$ & $R$ & $F_{1~\rm max}$ & $R$ & $F_{1~\rm max}$ & $R$    & $F_{1~\rm max}$ & $R$  & $F_{1~\rm max}$ & $R$     & Sec. & Ref. \\
   \hline
\multicolumn{15}{c}{Scatter-based indices} \\
 $\chi_{\rm red}^2$     & 0.166 & 0.927   & 0.184 & 0.936   & 0.873 & 0.991   & 0.556 & 0.990   & 0.563 & 0.991   & 0.685 & 0.993   & 0.252 & 0.980  & \ref{sec:chi2}         & (a) \\
 $\sigma_w$             & 0.179 & 0.948   & 0.186 & 0.941   & 0.860 & 0.992   & 0.493 & 0.990   & 0.577 & 0.991   & 0.669 & 0.993   & 0.258 & 0.980  & \ref{sec:sigma}        & (b) \\
 ${\rm MAD}$            & 0.180 & 0.958   & 0.264 & 0.966   & 0.825 & 0.992   & 0.548 & 0.993   & 0.561 & 0.991   & 0.779 & 0.992   & 0.400 & 0.989  & \ref{sec:MAD}          & (c) \\
 ${\rm IQR}$            & 0.175 & 0.954   & 0.247 & 0.962   & 0.826 & 0.992   & 0.556 & 0.993   & 0.608 & 0.992   & 0.777 & 0.992   & 0.384 & 0.988  & \ref{sec:IQR}          & (d) \\
 ${\rm RoMS}$           & 0.191 & 0.950   & 0.231 & 0.953   & 0.870 & 0.991   & 0.632 & 0.993   & 0.589 & 0.991   & 0.757 & 0.991   & 0.406 & 0.991  & \ref{sec:RoMS}         & (e) \\
 $\sigma_{\rm NXS}^2$   & 0.023 & 0.240   & 0.022 & 0.192   & 0.065 & 0.737   & 0.042 & 0.599   & 0.061 & 0.759   & 0.068 & 0.776   & 0.348 & 0.990  & \ref{sec:sigmaxs}      & (f) \\
 $v$                    & 0.096 & 0.898   & 0.107 & 0.911   & 0.668 & 0.993   & 0.400 & 0.989   & 0.471 & 0.991   & 0.165 & 0.957   & 0.096 & 0.952  & \ref{sec:peaktopeak}   & (g) \\
\multicolumn{15}{c}{Correlation-based indices} \\
 $l_1$                  & 0.708 & 0.993   & 0.545 & 0.993   & 0.877 & 0.991   & 0.864 & 0.992   & 0.759 & 0.991   & 0.887 & 0.992   & 0.689 & 0.993  & \ref{sec:lag1autocorr} & (h) \\
 $I$                    & 0.171 & 0.926   & 0.192 & 0.935   & 0.881 & 0.991   & 0.636 & 0.990   & 0.642 & 0.991   & 0.783 & 0.992   & 0.369 & 0.990  & \ref{sec:welch}        & (i) \\
 $J$                    & 0.211 & 0.953   & 0.242 & 0.957   & 0.884 & 0.991   & 0.756 & 0.993   & 0.653 & 0.991   & 0.780 & 0.991   & 0.419 & 0.994  & \ref{sec:stetson}      & (j) \\
 $J({\rm time})$        & 0.193 & 0.952   & 0.243 & 0.957   & 0.870 & 0.991   & 0.733 & 0.993   & 0.672 & 0.991   & 0.884 & 0.991   & 0.484 & 0.993  & \ref{sec:stetsonTime}  & (k) \\
 $J({\rm clip})$        & 0.198 & 0.949   & 0.232 & 0.953   & 0.892 & 0.991   & 0.704 & 0.992   & 0.613 & 0.992   & 0.768 & 0.991   & 0.403 & 0.993  & \ref{sec:stetsonClip}  & (d) \\
 $L$                    & 0.211 & 0.956   & 0.260 & 0.959   & 0.858 & 0.991   & 0.732 & 0.993   & 0.682 & 0.992   & 0.795 & 0.991   & 0.551 & 0.994  & \ref{sec:stetson}      & (j) \\
 ${\rm CSSD}$           & 0.197 & 0.950   & 0.324 & 0.974   & 0.018 & 0.009   & 0.019 & 0.014   & 0.019 & 0.007   & 0.019 & 0.010   & 0.019 & 0.001  & \ref{sec:cssd}         & (l) \\
 $E_x$                  & 0.509 & 0.988   & 0.413 & 0.989   & 0.854 & 0.992   & 0.700 & 0.993   & 0.622 & 0.992   & 0.784 & 0.992   & 0.430 & 0.994  & \ref{sec:excursions}   & (m) \\
 $1/\eta$               & 0.715 & 0.992   & 0.631 & 0.988   & 0.876 & 0.991   & 0.873 & 0.992   & 0.756 & 0.991   & 0.887 & 0.992   & 0.700 & 0.994  & \ref{sec:vonneumann}   & (n) \\
 $\mathcal{E}_\mathcal{A}$ & 0.257 & 0.978   & 0.405 & 0.990   & 0.828 & 0.991   & 0.692 & 0.994   & 0.480 & 0.991   & 0.885 & 0.991   & 0.525 & 0.994  & \ref{sec:EA}           & (o) \\
 $S_B$                  & 0.150 & 0.893   & 0.146 & 0.888   & 0.846 & 0.989   & 0.571 & 0.981   & 0.676 & 0.988   & 0.736 & 0.988   & 0.512 & 0.990  & \ref{sec:sb}           & (p) \\
   \hline
   \end{tabular}
   \renewcommand{\arraystretch}{1.0}
\vspace{-8pt}
\begin{flushleft}See the footnote in Table~\ref{tab:FRreal}.\end{flushleft}
\end{table*}

\clearpage

\end{document}